\documentclass[apj]{emulateapj}

\def\gtorder{\mathrel{\raise.3ex\hbox{$>$}\mkern-14mu
             \lower0.6ex\hbox{$\sim$}}}
\def\ltorder{\mathrel{\raise.3ex\hbox{$<$}\mkern-14mu
             \lower0.6ex\hbox{$\sim$}}}

\DeclareUnicodeCharacter{6768}{Yang}  
\DeclareUnicodeCharacter{8F76}{Yi}    
\usepackage{colordvi}
\usepackage{amsmath}
\usepackage{hyperref}
\usepackage{natbib}

\usepackage{lineno}

\def\gtorder{\mathrel{\raise.3ex\hbox{$>$}\mkern-14mu
             \lower0.6ex\hbox{$\sim$}}}
\def\ltorder{\mathrel{\raise.3ex\hbox{$<$}\mkern-14mu
             \lower0.6ex\hbox{$\sim$}}}

\slugcomment{Submitted to ApJ}
\shorttitle{AT\,2024wpp}

\begin{document}

\title{A search for minute-time-scale flares from the transient AT\,2024wpp}
\author{Eran~O.~Ofek\altaffilmark{1},
Lior~Ozer\altaffilmark{1},
Ruslan~Konno\altaffilmark{1},
Nimrod~Strasman\altaffilmark{1},
Ping~Chen\altaffilmark{1},
Sagi~Ben-Ami\altaffilmark{1},
David~Polishook\altaffilmark{1},
Alexander~Krassilchtchikov\altaffilmark{1},
Simone~Garrappa\altaffilmark{1},
Erez~A.~Zimmermann\altaffilmark{1},
Enrico~Segre\altaffilmark{1},
Asaf Horowicz\altaffilmark{1},
Avishay~Gal-Yam\altaffilmark{1},
Yarin M.~Shani\altaffilmark{1},
Stanislav Fainer\altaffilmark{1},
Michael Engel\altaffilmark{1},
Yahel Sofer-Rimalt\altaffilmark{1},
Anna~Y.~Q.~Ho\altaffilmark{2},
Yossi~Shvartzvald\altaffilmark{1},
Ofer~Yaron\altaffilmark{1},
Kris Rybicki\altaffilmark{1},
Arie~Blumenzweig\altaffilmark{1},
Sarah Spitzer\altaffilmark{1},
Ron Arad\altaffilmark{1}
}

\altaffiltext{1}{Department of Particle Physics and Astrophysics, Weizmann Institute of Science, 76100 Rehovot, Israel}
\altaffiltext{2}{Cornell University, Ithaca, NY 14853, USA}

\begin{abstract}

The AT\,2018cow-like fast blue optical transient AT\,2022tsd showed a large number of few-minute-duration, high-luminosity ($\sim10^{43}$\,erg\,s$^{-1}$) flares. 
We present an intensive search for such flares from another 18cow-like event, AT\,2024wpp. 
We have used the Large Array Survey Telescope (LAST) to observe this transient between 28 and 74 days after the approximate time of zero flux.
The target was observed for about 23\,hours 
to a sensitivity that allows one to detect $3\times10^{42}$\,erg\,s$^{-1}$ flares at $S/N\gtrsim5$.
No optical flares have been found, suggesting
a one-sided 2-$\sigma$ confidence upper limit of $<0.02$ on the flare's duty cycle, and flare rate lower than about $0.11$\,hr$^{-1}$.
These limits suggest that not all 18cow-like objects display a high rate of minute-timescale luminous flares. This can be explained either by diversity in the 18cow-like population or by viewing angle effects (e.g., beaming),
or rather that the optical depth towards the central emitting region did not fall below unity during the particular search window.
\end{abstract}


\section{Introduction}
\label{sec:Introduction}

Several lines of evidence link the formation of neutron stars (NS) and stellar-mass black holes (BH) 
to some types of stellar explosions, like core-collapse supernovae (SNe).
Although it is clear that these kinds of SNe are the result of the collapse of a massive star, the mechanism that expels the envelope,
and the nature of the leftover as a function of the initial conditions is still largely unclear (e.g., \citealt{Heger+2003ApJ_HowMassiveStarsEndLife_Supernovae_CoreCollapse}).
Furthermore, in the past 20\,years, a growing number of apparently new types of SN-related optical 
transients were found (e.g., \citealt{Quimby+2011_SLSN, 
Smith+2012_SN2010jp_PTF10aaxi_Halpha_triple,Drout+2014_Rapidly_Evolving,Arcavi+2016_FastTransients}).
An intriguing possibility is that some of these transients are powered by some sort of central engine (a remnant compact object), like a magnetar (\citealt{Thompson+2004ApJ_SN_MagnetarSpinDown_GRB, Woosley2010ApJ_BrightSN_Magnetar, Kasen+2016ApJ_MagnetarDrivenShockBreakout_SupernovaLightCurves}), an accreting BH (\citealt{Dexter+Kasen2013ApJ_SN_Powered_FallbackAcretion}), or a compact companion (e.g., \citealt{Chen+2024Natur12days_Period_Supernova}).

Recently, \cite{Ho+2023Natur_MinuteTimeScaleFlares_Transient} reported the discovery of 14 flares with a 10-min time scale,
associated with the extragalactic ($z=0.256$) transient AT\,2022tsd. 
These flares were detected as soon as one month and as late as four months after the transient maximum light.
During this period, the flare duty cycle was $\sim10\%$, and a few weeks later, the duty cycle dropped below about 3\%.
The flare luminosity ranged from $3\times10^{42}$\,erg\,s$^{-1}$ to about $10^{44}$\,erg\,s$^{-1}$, and their durations
ranged from about 10 to 80\,min.
Based on the fast variability ($\lesssim30$\,s) in the flare light curves, the emission size was constrained to $\lesssim9\times10^{11}\Gamma^{2}$\,cm,
where $\Gamma$ is the Lorentz factor in the emission region.
Combining this size limit and luminosity, the brightness temperature was estimated as $T_{\rm B}\gtrsim 2\times10^{10}\Gamma^{-4}$\,K (\citealt{Ho+2023Natur_MinuteTimeScaleFlares_Transient}).
This suggests that the emission is non-thermal and/or relativistic 
and may originate from some sort of compact object.
A key question is whether these flares exist in similar or other types of transients.
AT\,2022tsd belongs to the rare class of AT\,2018cow-like objects (\citealt{Prentice+2018ApJ2018cow_Discovery,Ho+2019_AT2018cow_Radio,Perley+2019_SN2018cow,Margutti+2019ApJ_AT2018cow_Xray,Ho+2023ApJ_SearchFBOT_ZTF_AT2018cow_Rate,Chen+2023ApJ_AT2018cow_HST_LatTimeUV}).
The main properties of this subgroup are: Time above half peak brightness of a few days; peak brightness similar to that of super-luminous supernovae; featureless spectra at early time with some indication for broad features; and relatively bright radio and X-ray emission. These properties were suggested to be the result of an interaction of a fast ejecta ($\gtrsim0.1$c) with circumstellar material (CSM) E.g.,
\cite{Ho+2019_AT2018cow_Radio, Margutti+2019ApJ_AT2018cow_Xray, Perley+2020_AT2020xnd_FastLuminousTransient, Nayana+Chandra2021ApJ_18cow_Radio, Ho+2022ApJ_18cow_2020xnd_Radio, Metzger+Perley2023ApJ_FBOT_18cow_DustEchos}.

The fact that the AT\,2022tsd-flares were serendipitously found in a rare class of transients may indicate either that these flares are associated only with 18cow-like events or that they are harder to detect in other kinds of SN events.
One simple possibility that we discuss in this paper is that the detection of such flares requires seeing through the SN ejecta, and this is possible only when the optical depth to the explosion site drops below unity.
The time scale for which the optical depth between the observer and any compact remnant becomes smaller than unity, 
depends on the ejecta mass, velocity, and opacity.
For the majority of regular core-collapse SNe, with ejecta mass of a few M$_{\odot}$, this time scale is of the order of a year.
This may make the flaring events hard to detect.

Finding additional examples, among different classes of SNe, is of significant importance. 
First, it would shed light on the origin of the flares.
Second, it may provide some important clues on the nature of the remnants left over in different kinds of SNe.

In this paper, we present a targeted search using the Large Array Survey Telescope (LAST; \citealt{Ofek+2023PASP_LAST_Overview, BenAmi+2023PASP_LAST_Science}),
for minute-timescale flares from the 18cow-like transient AT\,2024wpp.
AT\,2024wpp was discovered on 2024 Sep 26 by the Zwicky Transient Facility (ZTF; \citealt{Bellm+2019_ZTF_Overview, Graham+2019_ZTF_objectives, Masci+2019_ZTF_Pipeline}).
Pre-discovery observations suggested it has a relatively fast rise (\citealt{Ho+2024TNSAN_AT2024wpp_discovery}).
The object's redshift was measured to be $0.0868$ (\citealt{Sfaradi+2024TNSAN_AT2024wpp_redshift, Perley+2024TNSAN_AT2024wpp_Redshift}).
Radio observations (\citealt{Schroeder+2024TNSAN_AT2024wpp_VLA}) using the Very Large Array (VLA), obtained about a month after the discovery, identified a point source with
flux density of $254\,\mu$Jy in X-band ($10$\,GHz)
and $311\,\mu$Jy in Ku-band ($15$\,GHz).
This corresponds to a $10$\,GHz luminosity of about $5\times10^{28}$\,erg\,s$^{-1}$\,Hz$^{-1}$ at 29 days post-discovery in the observer frame. This is a factor of a few higher than the radio luminosity of AT\,2018cow at a similar frequency and epoch (\citealt{Margutti+2019ApJ_AT2018cow_Xray}).
It was also detected in the X-rays, three days after the first optical detection, using {\it Swift}-XRT (\citealt{Gehrels+2004_Swift}), with an unabsorbed luminosity of about $2\times10^{43}$\,erg\,s$^{-1}$
(\citealt{Srinivasaragavan+2024TNSAN_AT2024wpp_Xray}). %
\cite{Margutti+2024TNSAN_AT2024wpp_Xray}
reported on additional X-ray observations using {\it Swift}-XRT and {\it Chandra}, where they found a soft X-ray source with a roughly constant flux until $\sim7$\,days after the discovery, followed by a flux decay.
At around $\sim50$\,days after discovery, the source experienced a rapid X-ray brightening over a time scale of a few days, and a significant spectral hardening.
\cite{Pursiainen+2025MNRAS_AT2024wpp_spherical} reported, based on a black-body fit as a function of time, that the photosphere of AT\,2024wpp was rapidly expanding at $\sim1.2\times10^{4}$\,km\,s$^{-1}$.
Furthermore, the spectra of this event were consistent with a blackbody,
with a tentative broad feature at $\sim5500$\,\AA.~They also reported on optical polarization $\lesssim0.5$\% between 6 to 14 days relative to maximum light.
On the other hand, \cite{Maund+2023MNRAS_AT2018cow_polarization} reported polarimetric observations of AT\,2018cow. They measure a polarization
of about 7\% (in $r$-band) around 6\,days following explosion, dropping fast to the 1\% level a day later, and spiking again at 2\% (in $B$-band), around 2 weeks after explosion.
Most of these observational features of AT\,2024wpp show resemblance to a 18\,cow-like fast transient (\citealt{Ho+2019_AT2018cow_Radio, Margutti+2019ApJ_AT2018cow_Xray, Perley+2019_SN2018cow}).

Here, our main objective is to constrain the flaring activity in AT\,2024wpp.
We analyze about 155 observing hours conducted using LAST,
with a typical limiting magnitude ranging between 20 and 21.5 mag.
Of the total observing time, about $\approx35$ hours are non-overlapping (in 6-min time resolution).
We did not find any flares associated with AT\,2024wpp, and 
we set a limit on the duty cycle and flare rate as a function of flare absolute magnitude.
The non-detection may suggest diversity in physical properties of 18\,cow-like transients.

In \S\ref{sec:considerations} we discuss some theoretical considerations that guide our search strategy.
In \S\ref{sec:obs} we discuss the 
observations and data reduction,
while the data analysis is presented in \S\ref{sec:analysis}, and we conclude in \S\ref{sec:conc}.

\section{Optical-depth consideration}
\label{sec:considerations}

The main purpose of this work is to look for minute-time-scale flares from another 18\,cow-like event.
Since we assume, based on the evidence we have so far, that the flares take place at the explosion site, a requirement for the detection of these flares is that the optical depth to the explosion site will be smaller than unity.
This optical depth
is given by:
\begin{equation}
    \tau = \int{\kappa \rho dr} \sim 2 \frac{\kappa}{0.3\,{\rm cm}^{2}\,{\rm g}^{-1}} \frac{M_{\rm ej}}{{\rm M}_{\odot}} \Big( \frac{v}{10^{4}\,{\rm km\,s}^{-1}} \frac{t}{100\,{\rm d}} \Big)^{-2}.
    \label{eq:OpticalDepth}
\end{equation}
Here, $\kappa$ is the opacity, $M_{\rm ej}$ is the ejecta mass, $v$ is the ejecta velocity, $\rho$ is the ejecta density, $r$ is the radius of the ejecta, and $t$ is the time after the explosion.
Since the opacity, ejecta density profile, and mass are poorly constrained, this is an order of magnitude estimate for $\tau$.

Due to the photon diffusion, $\tau$ even slightly above unity, can smear the light curve of the flare.
Specifically, the photon diffusion time scale is $t_{\rm diff}\sim r^{2}\rho\kappa/c\approx r\tau/c$.
For $\tau=1$, and $r\gtrsim2\times10^{13}$\,cm, $t_{\rm diff}\gtrsim600$\,s.
Here, we choose a radius lower limit which is of the order of the photosphere size at $\lesssim1$\,day.
Therefore, $\tau<1$ is an essential requirement.
In Equation~\ref{eq:OpticalDepth}, we assume that the opacity is due to Thomson scattering. In principle, if some of the gas had enough time to cool down, the effective opacity may be lower.

Therefore, the fact that flares from AT\,2022tsd were detected about one month after the explosion can be possibly explained by:
(i) aspherical explosion in which we have a line-of-sight with low $\tau$;
(ii) ejecta mass $\lesssim1$\,M$_{\odot}$.
The low ejecta mass possibility is consistent with the fast time evolution of AT\,2022tsd, which suggests a short photon diffusion time scale and, hence, a low ejecta mass.

We can not rule out that the ejecta and/or circumstellar material (CSM) in this transient
are a-spherical. \cite{Margutti+2019ApJ_AT2018cow_Xray}
suggested that a-sphericity can explain
some of the complicated features of AT\,2018cow
(e.g., late time appearance of low-velocity [$4000$\,km\,s$^{-1}$] Balmer emission lines,
spectral slope, and early-time broad features).
Their model suggests that we were observing AT\,2018cow from a near-polar direction, which was dominated by low-mass, fast ($\sim0.1$c), ejecta.
This is consistent with the low optical depth,
at about 2 months after the explosion,
required to explain the detectability of flares from a central engine.

\section{Observations and data reduction}
\label{sec:obs}

\begin{figure}
\centerline{\includegraphics[width=8cm]{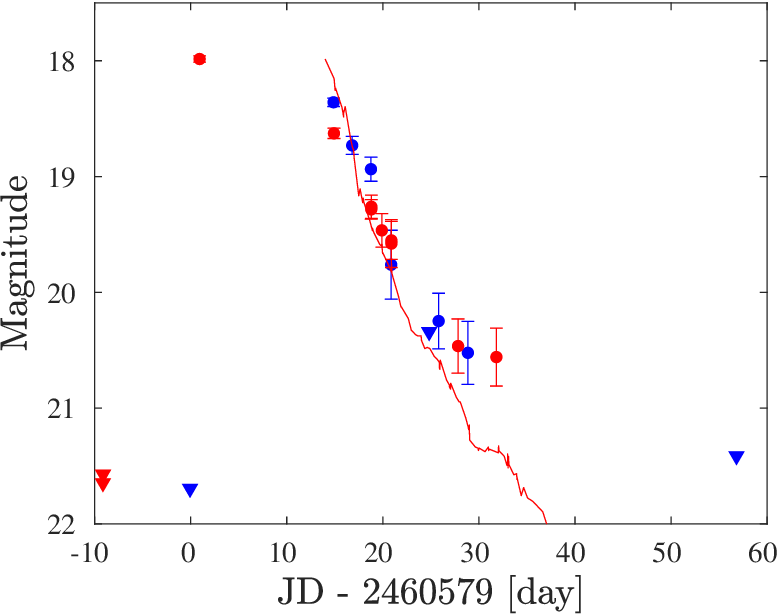}}
\caption{The ZTF $g$ (blue circles) and $r$ (red circles) light curve of AT\,2024wpp. Time is measured relative to JD~2460579,
which is the approximate time of zero flux. The triangles mark $3$-$\sigma$ upper limits on the magnitude.
The red line shows the $r$-band light curve of AT\,2018cow (adopted from \citealt{Perley+2019_SN2018cow}). The LC is scaled to the distance modulus of AT\,2024wpp (added 4.0558\,mag) and the time is relative to JD $2458301.7674$.
\label{fig:AT2024wpp_ZTF_LC}}
\end{figure}

In Figure~\ref{fig:AT2024wpp_ZTF_LC}, we present the optical light curve of AT\,2024wpp as obtained using the ZTF survey (\citealt{Bellm+2019_ZTF_Overview}).
The light curve was obtained from the forced photometry service (\citealt{Masci+2019_ZTF_Pipeline}).
See also photometric light curve in \cite{Pursiainen+2025MNRAS_AT2024wpp_spherical}.
Here, we define the time of zero-flux as Julian Day (JD)$=2460579$, which corresponds to the rounded JD time between the first detection and the previous non-detection.

We observed AT\,2024wpp using the Large Array Survey Telescope (LAST) system.
LAST is an array currently consisting of 40 telescopes, which will increase to 
72 telescopes at the end of phase I.
Groups of four telescopes are mounted on a single mount, which provides great flexibility in observing strategy.
Each telescope is a 28-cm f/2.2 Rowe-Ackermann Schmidt with a field of view of about 7.4\,deg$^{2}$, and a pixel scale of 1.25\,arcsec\,pix$^{-1}$.
The default LAST strategy is that $20\times20$\,s images are obtained for each field (with zero dead time between the images).
This set of $20\times20$\,s images is called a {\it visit}.
The real-time image reduction pipeline is presented in \cite{Ofek+2023PASP_LAST_PipeplineI}, while
the image subtraction and transient detection are discussed in Konno et al. (in prep).
Under dark conditions\footnote{The $V$-band sky brightness at the LAST site is about 20.8\,mag\,arcsec$^{-2}$.}, a single LAST telescope reaches a limiting magnitude of about 20.8 (19.5) in a $20\times20$\,s (20\,s) image.
However, due to frequent out-of-focus problems during the commissioning period, the actual limiting magnitude distribution is broad.
LAST is further described in 
\cite{Ofek+BenAmi2020_Grasp_SkySurvrys_CostEffectivness}, \cite{Ofek+2023PASP_LAST_Overview},
and \cite{BenAmi+2023PASP_LAST_Science}.

\begin{figure}
\centerline{\includegraphics[width=8cm]{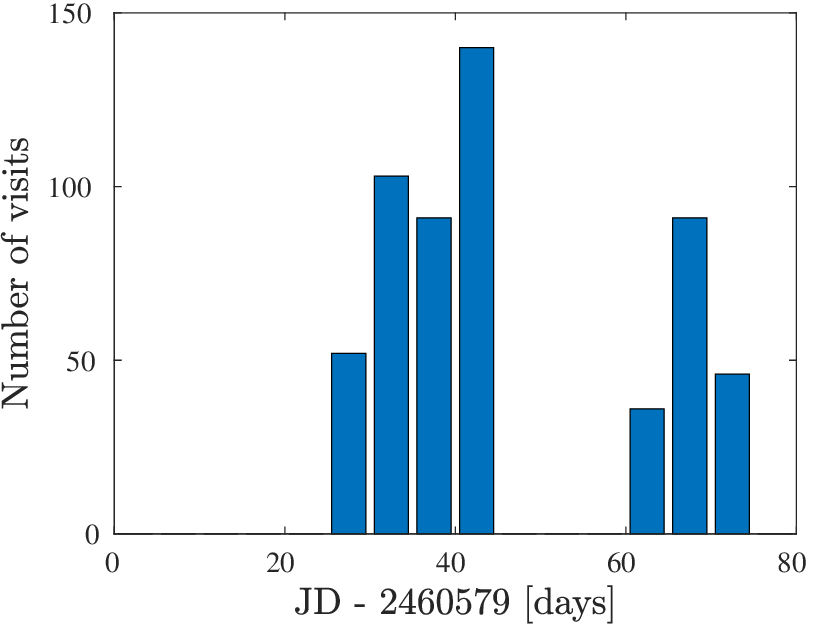}}
\caption{Histogram of the number of LAST visits (a sequence of $20\times20$\,s exposures) of AT\,2024wpp,
as a function of the time relative to the transient approximate zero-flux date (JD$=2460579$). Bin size is 5\,days. This includes visits that were taken by different telescopes at the same time.
\label{fig:AT2024wpp_Visit_Time}}
\end{figure}

We have used LAST to observe AT\,2024wpp starting 2024 Oct 23, about a month after the transient discovery, until 2024 Dec 8, about 74\,days after the approximate time of zero flux.
The number of visits as a function of time after the approximate time of zero flux is shown in Figure~\ref{fig:AT2024wpp_Visit_Time}.
Some of the data were obtained using telescopes that were observing the same sky position.
This reduces the amount of effective time we spend on target. On the other hand, it allows us to go deeper and gives us yet another tool to verify the existence of flares, if detected.

Our data analysis is based on image subtraction of the standard LAST visit image 
(coadded from $20\times20$\,s individual images).
Our reference images (per telescope) were created from the same data, typically from the 50 images with the best seeing.
The coaddition was done using weighted mean sigma clipping.
The sigma-clipping means that if there is a flare within the data used to generate the reference image, it is likely to be removed. However, this means that if the duty cycle of the flares is higher than about 20--30\%, then the flares may contaminate the reference image. In such a case, the flares can still be detected, but with a lesser sensitivity.
Here, the weights are given by transmission divided by the image background variance (e.g., \citealt{Zackay+2017_CoadditionI}),
where the transmission is proportional to $10^{0.4\,{\rm ZP}}$, where ${\rm ZP}$ is the photometric zero point of the image in magnitudes.
The ZP of the new and reference images was estimated by calibrating the LAST instrumental magnitude to GAIA-DR3 (\citealt{GAIA+2016_GAIA_mission, GAIA+2022yCat_GAIA_DR3_MainSourcesCatalog})
$B_{\rm p}$ magnitude, including a $B_{\rm p}-R_{\rm p}$ color term.
The zero point is evaluated assuming the source's color is $B_{\rm p}-R_{\rm p}=1$\,mag
(see \citealt{Ofek+2023PASP_LAST_PipeplineI}).

\begin{figure*}
\centerline{\includegraphics[width=15cm]{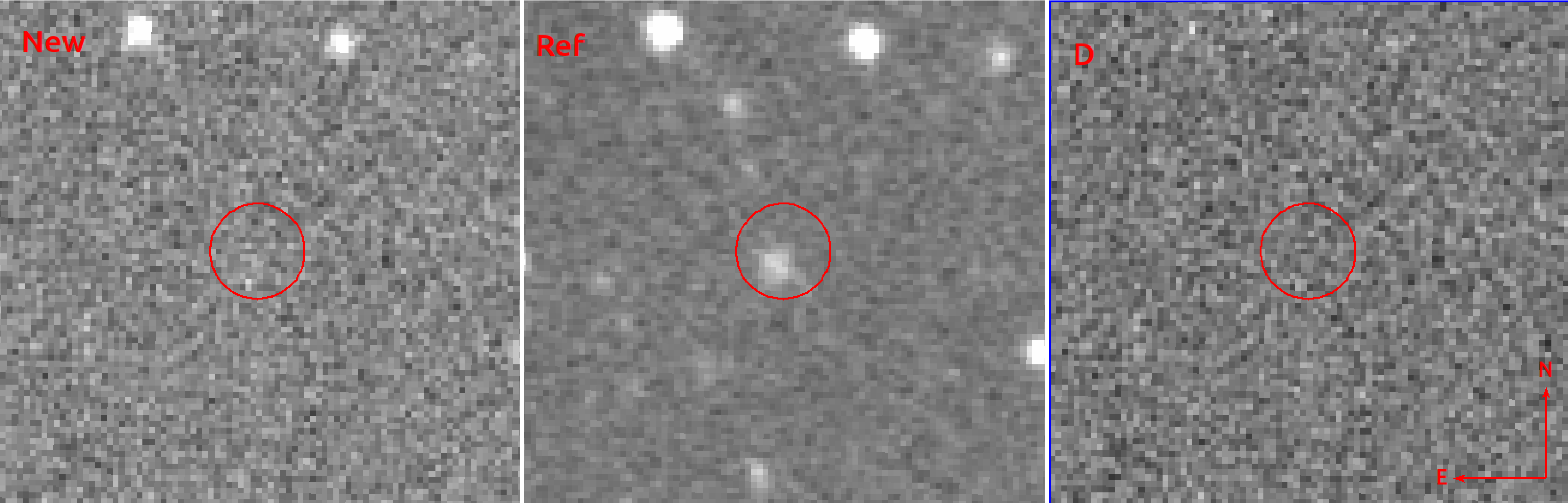}}
\caption{Example of New, Reference, and subtraction ($D$) images (left to right) of AT\,2024wpp. The transient position is marked with a 10\,arcsec-radius red circle.
\label{fig:SN2024wpp_RND_example}}
\end{figure*}

The image subtraction was done using the \cite{Zackay+2016_ZOGY_ImageSubtraction} (ZOGY) algorithm with additional metadata calculated using the {\it Translient} algorithm (\citealt{Springer+Ofek+2024AJ_Translient}).
The translient statistics ($Z^{2}$) is a simple hypothesis testing between the null hypothesis that the source is stationary and the alternative hypothesis that the source is moving. Therefore, comparison of $Z^{2}$ with $S^{2}$ indicates if the subtraction residuals are due to motion (e.g., atmospheric scintillation), or due to variability.
The flux matching is done by the images' photometric zero point.
The flux in the transient location is determined using PSF-fitting of the image subtraction PSF to the proper subtraction image.
An example of a triplet of reference, new and subtraction ($D$) images is shown in Figure~\ref{fig:SN2024wpp_RND_example}.

The fact that the images we analyze are based on a coadd of 20 images has several important advantages.
These coadded images are much cleaner, with substantially reduced numbers
of cosmic ray, streaks, and bad pixels affecting their quality.
Furthermore, this strategy filters out
satellite glints, which have a high rate (e.g., \citealt{Corbett+2020_SatellitesGlints, Nir+2020_Satellites_Glints_FlaresLimit, Nir+2021_RNASS_GN-z11-Flash_SatelliteGlint}).
Finally, it allows us to check the reality (and inspect the temporal evolution) of any detected event using the individual images.
Given this search strategy, we are sensitive to events that are longer than about 5\,min.
Given the typical duration of flares observed in AT\,2022tsd ($\sim10-80$\, min), this timescale is of interest.

In each visit, we calculate the 5~$\sigma$ limiting magnitude
of the difference image, as follows:
In the ZOGY scheme, we can calculate the parameter $F_{\rm D}$ (see Equation~15 in ZOGY), and hence convert the zero point of the new image to the zero point of the difference image.
Next, we measure the standard deviation of the difference image ($D$ in ZOGY notation),
and multiply it by the number of standard deviations we require for detection (i.e., 5), which we refer to as the per-pixel flux limit.
Since the detection process in ZOGY is equivalent to a matched filter, we have to correct the per-pixel flux limit by dividing it by $\sqrt{\sum{P_{\rm D}^2}}$, where $P_{\rm D}$ is the PSF of the difference image and the summation is over all the pixels of the PSF (e.g., \citealt{Zackay+2017_CoadditionII}).
Finally, we can convert this to magnitude by using the zero point of the difference image.
This scheme is accurate when we are in the background-noise-dominated regime.
In our case, since the host galaxy (and any transient light) is fainter than the sky, this assumption is justified.
%
In Table~\ref{tab:Obs}, we provide all the measurements resulting from this process.
\begin{deluxetable*}{llrllcclrlllll}
\tablecolumns{14}
\tablewidth{0pt}
\tablecaption{LAST observations of AT2024wpp}
\tablehead{
\colhead{JD}    &
\colhead{$S$}   &
\colhead{$Z^{2}$} &
\colhead{$m_{\rm lim}$} &
\colhead{Mount} &
\colhead{CamNum} &
\colhead{CropID} &
\colhead{FLAGS} &
\colhead{PSF Flux} &
\colhead{SN} &
\colhead{$\chi^{2}$} &
\colhead{ZP} &
\colhead{Bck} &
\colhead{Bck Std}\\
\colhead{(day)}       &
\colhead{()}       &
\colhead{()}       &
\colhead{(mag)}    &
\colhead{()} &
\colhead{()} &
\colhead{()} &
\colhead{()} &
\colhead{(cnt)} &
\colhead{()} &
\colhead{()} &
\colhead{(mag)} &
\colhead{(cnt)} &
\colhead{(cnt)}
}
\startdata
$2460642.29440$  &  $-0.349$  &  $ 0.641$  &  $20.42$  &  $ 4$  &  $1$  &  $16$  &  $  746586252$  &  $  48.64$  &  $  6.50$  &  $ 20.600$  &  $25.49$  &  $-0.324$  &  $ 3.108$  \\ 
      $2460642.30297$  &  $-0.024$  &  $ 0.504$  &  $20.47$  &  $ 4$  &  $1$  &  $16$  &  $  746586244$  &  $  28.58$  &  $  4.73$  &  $ 24.578$  &  $25.49$  &  $-0.486$  &  $ 2.882$  \\ 
      $2460642.30760$  &  $ 0.917$  &  $ 0.560$  &  $20.42$  &  $ 4$  &  $1$  &  $16$  &  $  746586252$  &  $  19.93$  &  $  3.49$  &  $ 13.648$  &  $25.48$  &  $-0.701$  &  $ 2.715$  \\ 
      $2460642.31223$  &  $-0.665$  &  $ 3.684$  &  $20.36$  &  $ 4$  &  $1$  &  $16$  &  $  746586252$  &  $  46.59$  &  $  6.35$  &  $ 51.780$  &  $25.48$  &  $-0.519$  &  $ 2.690$  \\ 
      $2460642.31797$  &  $ 0.067$  &  $ 2.291$  &  $20.34$  &  $ 4$  &  $1$  &  $16$  &  $  746586252$  &  $   9.50$  &  $  2.13$  &  $ 11.064$  &  $25.46$  &  $-0.576$  &  $ 2.793$ 
\enddata
\tablecomments{LAST photometric visit measurements of AT\,2024wpp. JD is Julian days; $S$ is the ZOGY image subtraction detection statisics (\citealt{Zackay+2016_ZOGY_ImageSubtraction}); $Z^{2}$ is the {\it Translient} motion detector statistics (\citealt{Springer+Ofek+2024AJ_Translient}); 
$m_{\rm lim}$ is the 5-$\sigma$ limiting magnitude in the subtraction image (see text);
Mount - LAST mount number;
CamNum - LAST camera number on mount;
CropID - LAST sub image crop ID;
FLAGS - Bit mask flags at the photometric position (see \citealt{Ofek+2023PASP_LAST_PipeplineI});
PSF flux - The PSF fitted flux;
SN is the $S/N$ ratio of the PSF fitting;
$\chi^{2}$ is the total $\chi^{2}$ of the PSF fitting (with about 28 degrees of freedom);
ZP is the photometric zero point that is used to convert the flux to magnitude;
Bck is the background in an annulus around the transient location;
and Bck Std is the std of the background in an annulus around the transient location.
}
\label{tab:Obs}
\end{deluxetable*}

\section{Analysis}
\label{sec:analysis}

\begin{figure}
\centerline{\includegraphics[width=8cm]{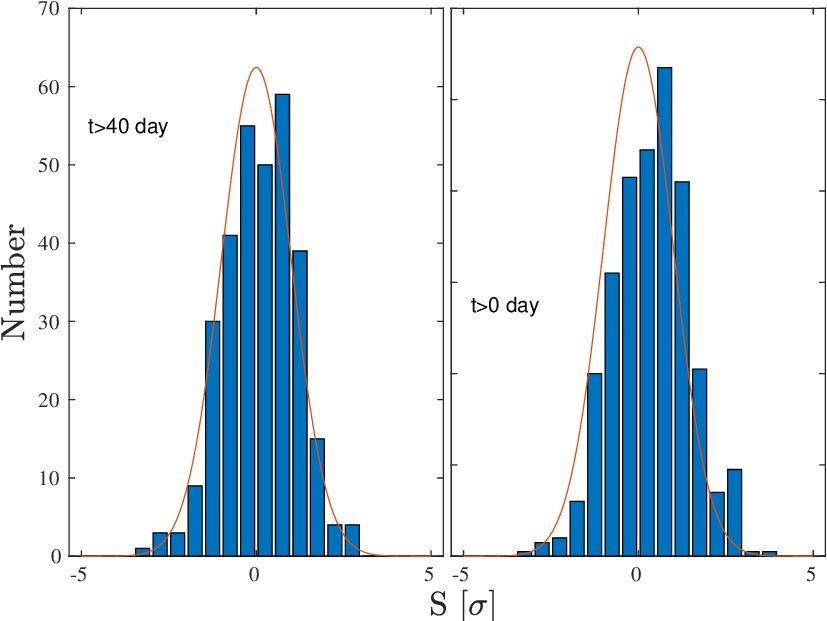}}
\caption{The distribution of the ZOGY transient detection statistics $S$ of the image subtraction process (in units of Gaussian $\sigma$).
The red curves present the theoretical expectation of a Gaussian with zero mean and unit standard deviation.
The right panel shows the distribution for all the data points,
while the left panel is for data points obtained more than 40 days after JD $2460579.0$. See text for discussion.
\label{fig:AT2024wpp_Hist_S}}
\end{figure}

Since in ZOGY the primary detection criterion is the detection score $S$ (in units of Gaussian noise $\sigma$; Equation 17 in ZOGY), we perform the detection via searching for events with $S>5$.
Figure~\ref{fig:AT2024wpp_Hist_S} presents the histogram of $S$ over all the visits we obtained.
The red curve shows the expected Gaussian distribution of the noise.
The right panel presents the histogram for all the data.
In this case we see that the measurements slightly deviate from the expected Gaussian curve towards positive values.
Since the reference was constructed from a mix of all the images, one possibility is that this is due to some AT\,2024wpp light leftover in some of the new images.
If this hypothesis is correct, then we expect it to be more prominent at early times after the explosion.
Therefore, in the left panel, we show the distribution of $S$ only for late times (more than 40 days after the approximate zero-flux time). Indeed, in this case, the distribution of $S$ is consistent with the Gaussian expectation.

\begin{figure}
\centerline{\includegraphics[width=8cm]{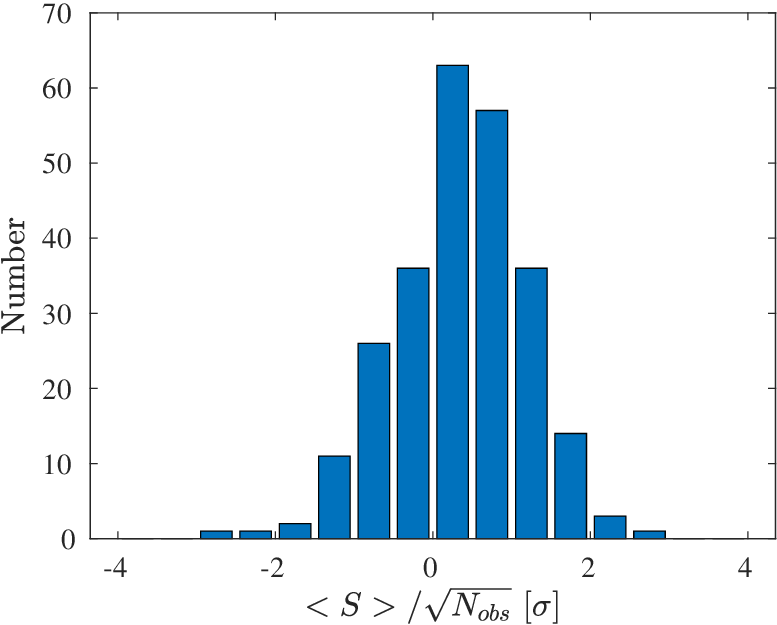}}
\caption{Histogram of the renomalized mean $S$ values in 10-min time bins. 
\label{fig:AT2024wpp_MeanS}}
\end{figure}

Since we are interested in detecting flares with a time scale of $\gtrsim10$\,min, and since in some cases we have multiple telescopes observing the field at the same time, we binned the data into 10-min bins, where the first bin window started 5\,min prior to the first observation.
In each 10-minute bin, we count the number of data points
($N_{\rm obs}$; visits), and the mean value of $S$ in the bin.
In each bin, we renormalize the mean $S$ by dividing it by
$\sqrt{N_{\rm obs}}$ so that the standard deviation of the renormalized $S$ will be approximately one.
The mean (median) $N_{\rm obs}$ is about 1.6 (2).
In Figure~\ref{fig:AT2024wpp_MeanS} we present the histogram of renormalized mean $S$ values in 10-min time bins.
The main result is that we do not detect any flaring event with time scale $\gtrsim10$\,min, above 3$\sigma$ detection limit.

\begin{figure}
\centerline{\includegraphics[width=8cm]{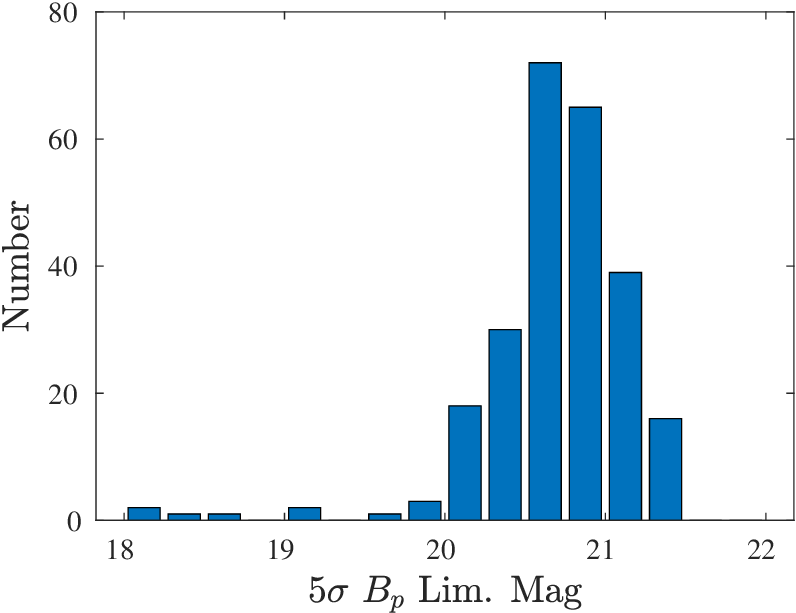}}
\caption{Histogram of the $5\sigma$ $B_{\rm p}$ limiting magnitude in the 10-min time bins. 
\label{fig:AT2024wpp_LimMag}}
\end{figure}

We also calculated the mean limiting magnitude of all the visits in the bin ($m_{\rm lim}$).
We use this to calculate the approximate combined limiting magnitude in the bin: $m_{\rm lim}+1.25\log_{10}{N_{\rm obs}}$.
This approximation is valid if the spread of the limiting magnitudes within each bin is small.
Indeed, 90\% (95\%) of the bins have a range of limiting magnitude smaller than $0.20$ ($0.26$)\,mag.
Figure~\ref{fig:AT2024wpp_LimMag} presents the histogram of the 5~$\sigma$ limiting magnitude in the 10-min bins.

\begin{figure}
\centerline{\includegraphics[width=8cm]{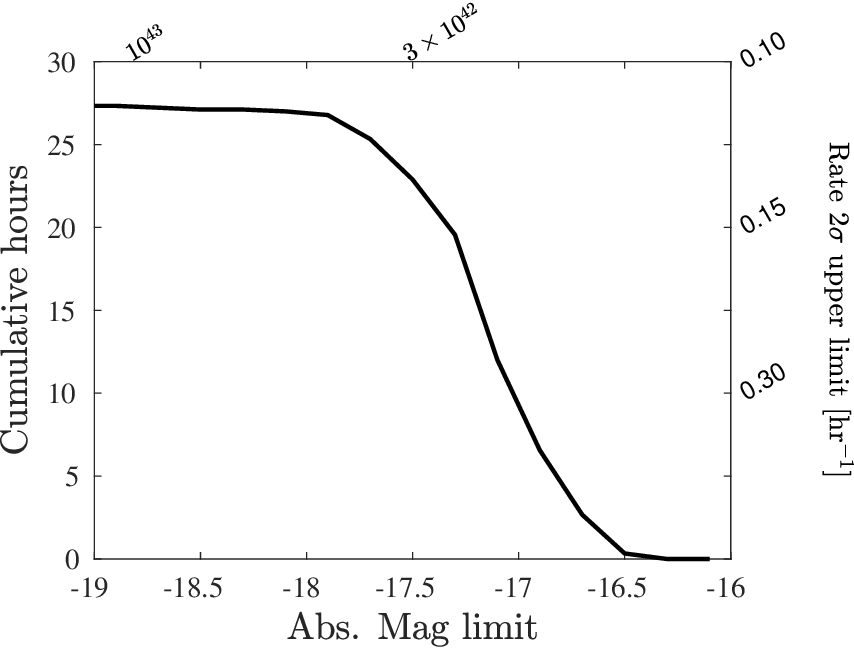}}
\caption{The cumulative independent observing time of AT\,2024wpp, as a function of the absolute magnitude corresponding to the 5~$\sigma$ detection limit.
The upper x-axis indicates the approximate luminosity, in erg\,s$^{-1}$, corresponding to the absolute magnitude.
\label{fig:AT2024wpp_CumHours_LimAbsMag}}
\end{figure}

Finally, we convert the apparent limiting magnitude to absolute magnitudes using the source redshift $z=0.0868$ and WMAP cosmology (\citealt{Hinshaw+2013_WMAP_9yr_CosmologicalParameters}; distance modulus of $37.979$\,mag.),
and calculate the cumulative amount of observing time that can detect an event brighter than some absolute limiting magnitude.
Figure~\ref{fig:AT2024wpp_CumHours_LimAbsMag} presents the cumulative number of independent hours in which we were sensitive to 10-min flare brighter than some absolute magnitude.
In the upper X-axis, we provide the corresponding luminosity in erg\,s$^{-1}$.
The luminosity was calculated assuming a solar-like spectrum, and Sun's absolute magnitude of 4.7\,mag.
On the right side Y-axis, we also provide the 2-$\sigma$ upper limit on the rate in units of hr$^{-1}$ (corresponding to the number of observing hours).
This was calculated by $3.0$ divided by the number of observing hours, where $3.0$ is the one-sided $2\sigma$ Poisson upper limit for zero detections (e.g., \citealt{Gehrels1986_PoissonCI}).
We note that in the case that the flare's time distribution
is highly correlated (as seen in e.g., soft gamma-ray repeaters; 
\citealt{Gogus+1999ApJ_SGR_BurstDistributionStatistics}), then these limits may be less constraining.
The number of observing hours can be roughly translated to a one-sided $2\sigma$ upper limit on the duty cycle
by $3.0\Delta{t}/(60 T_{\rm obs})$,
where $\Delta{t}$ is the assumed duration of the flare [minutes], and $T_{\rm obs}$ is the total number of observing hours.
The reason the formula of the duty cycle, from observing hours, is an approximation (even for a pure Poisson process) is that it neglects to take into account the edges of observing blocks that may slightly increase the effectiveness of the search.

\section{Conclusions}
\label{sec:conc}

We present a search for minutes-time-scale flares from AT\,2024wpp, using LAST. We did~not find any such flares in the time window between 28 to 74 days after the transient discovery.
We set a one-sided $2\sigma$ upper limit on the flare rate (duty cycle) of $0.11$\,hr$^{-1}$ ($<0.02$).
These limits are calculated for 10-min duration flares brighter than $3\times10^{42}$\,erg\,s$^{-1}$ 
(similar to the faintest flares observed from AT\,2022tsd).

To put these limits in context,
so far, such minute-time-scale flares were detected only in the case of AT\,2022tsd (\citealt{Ho+2023Natur_MinuteTimeScaleFlares_Transient}).
In the case of AT\,2022tsd, between one month and four months after the approximate explosion time, 
the duty cycle of the flares was $\sim10\%$,
and each flare had a typical luminosity of the order of $3\times10^{42}-10^{44}$\,erg\,s$^{-1}$.
During this period, the flare rate was about $0.29_{-0.08}^{+0.10}$\,hr$^{-1}$.
Furthermore, in an intensive search done about 4.5\,months after the explosion time, no flares were found, suggesting that the activity significantly decreased, or stopped\footnote{This can also happen if the activity is temporally correlated.}, at some late times.
%
We conclude that the minute-time-scale flare activity level of AT\,2024wpp was likely significantly lower compared to that of AT\,2022tsd.

%
There are several possibilities explaining the lack of flares from this object:
(i) There are sub-classes among the 18-cow-like events, and only some of them demonstrate minute-scale flares;
(ii) The flares are visible only from specific viewing angles. This can happen if the ejecta is a-spherical and, hence, optically thin only in certain directions,
or if the emission originates from a relativistic jet -- although AT\,2018cow does show a strong polarization signal (\citealt{Maund+2023MNRAS_AT2018cow_polarization}), while the polarization measurements of AT\,2024wpp do not show evidence for strong polarization (\citealt{Pursiainen+2025MNRAS_AT2024wpp_spherical}), it is worthwhile to obtain more polarimetric observations of 18-cow-like transients.
A-sphericity can also explain the subtle differences (e.g., X-ray luminosity) between the various 18cow-like objects (\citealt{Margutti+2019ApJ_AT2018cow_Xray});
(iii) The ejecta mass in AT\,2024wpp is considerably larger or has a different distribution compared to that of AT\,2022tsd, and therefore, at the time we observed the target, it was optically thick;
or (iv) the flares are highly temporally correlated.

Given the fact that the observations and characterization of such minutes-time scale flares from all kinds of transients and supernovae can give us important clues regarding possible central engines left over after the explosion, we have initiated two such searches using the data from the LAST telescope array.
The first is a targeted search among known supernovae, and the second is a blind search for minute time-scale flares, regardless of known supernovae. 

\acknowledgments

We would like to thank an anonymous referee for useful comments on the manuscript.
E.O.O. is grateful for the support of
grants from the 
Willner Family Leadership Institute,
André Deloro Institute,
Paul and Tina Gardner,
The Norman E Alexander Family M Foundation ULTRASAT Data Center Fund,
Israel Science Foundation,
Israeli Ministry of Science,
Minerva,
BSF, BSF-transformative, NSF-BSF,
Israel Council for Higher Education (VATAT),
Sagol Weizmann-MIT,
Yeda-Sela, and the
Rosa and Emilio Segr\`e Research Award.
This research is supported by the Israeli Council for Higher Education (CHE) via the Weizmann Data Science Research Center, and by a research grant from the Estate of Harry Schutzman.
This work is based on data collected using the Large Array Survey Telescope (LAST).
LAST is operated by the Weizmann Institute of Science.
This work uses data obtained using the Zwicky Transient Facility (ZTF; 
\citealt{ZTF_IPAC_LightCurves}).

\bibliography{papers.bib}

\begin{thebibliography}{50}%
\makeatletter
\providecommand \@ifxundefined [1]{%
 \@ifx{#1\undefined}
}%
\providecommand \@ifnum [1]{%
 \ifnum #1\expandafter \@firstoftwo
 \else \expandafter \@secondoftwo
 \fi
}%
\providecommand \@ifx [1]{%
 \ifx #1\expandafter \@firstoftwo
 \else \expandafter \@secondoftwo
 \fi
}%
\providecommand \natexlab [1]{#1}%
\providecommand \enquote  [1]{``#1''}%
\providecommand \bibnamefont  [1]{#1}%
\providecommand \bibfnamefont [1]{#1}%
\providecommand \citenamefont [1]{#1}%
\providecommand \href@noop [0]{\@secondoftwo}%
\providecommand \href [0]{\begingroup \@sanitize@url \@href}%
\providecommand \@href[1]{\@@startlink{#1}\@@href}%
\providecommand \@@href[1]{\endgroup#1\@@endlink}%
\providecommand \@sanitize@url [0]{\catcode `\\12\catcode `\$12\catcode `\&12\catcode `\#12\catcode `\^12\catcode `\_12\catcode `\%12\relax}%
\providecommand \@@startlink[1]{}%
\providecommand \@@endlink[0]{}%
\providecommand \url  [0]{\begingroup\@sanitize@url \@url }%
\providecommand \@url [1]{\endgroup\@href {#1}{\urlprefix }}%
\providecommand \urlprefix  [0]{URL }%
\providecommand \Eprint [0]{\href }%
\providecommand \doibase [0]{http://dx.doi.org/}%
\providecommand \selectlanguage [0]{\@gobble}%
\providecommand \bibinfo  [0]{\@secondoftwo}%
\providecommand \bibfield  [0]{\@secondoftwo}%
\providecommand \translation [1]{[#1]}%
\providecommand \BibitemOpen [0]{}%
\providecommand \bibitemStop [0]{}%
\providecommand \bibitemNoStop [0]{.\EOS\space}%
\providecommand \EOS [0]{\spacefactor3000\relax}%
\providecommand \BibitemShut  [1]{\csname bibitem#1\endcsname}%
\let\auto@bib@innerbib\@empty
\bibitem [{\citenamefont {{Heger}}\ \emph {et~al.}(2003)\citenamefont {{Heger}}, \citenamefont {{Fryer}}, \citenamefont {{Woosley}}, \citenamefont {{Langer}},\ and\ \citenamefont {{Hartmann}}}]{Heger+2003ApJ_HowMassiveStarsEndLife_Supernovae_CoreCollapse}%
  \BibitemOpen
  \bibfield  {author} {\bibinfo {author} {\bibfnamefont {A.}~\bibnamefont {{Heger}}}, \bibinfo {author} {\bibfnamefont {C.~L.}\ \bibnamefont {{Fryer}}}, \bibinfo {author} {\bibfnamefont {S.~E.}\ \bibnamefont {{Woosley}}}, \bibinfo {author} {\bibfnamefont {N.}~\bibnamefont {{Langer}}}, \ and\ \bibinfo {author} {\bibfnamefont {D.~H.}\ \bibnamefont {{Hartmann}}},\ }\href {\doibase 10.1086/375341} {\bibfield  {journal} {\bibinfo  {journal} {\apj}\ }\textbf {\bibinfo {volume} {591}},\ \bibinfo {pages} {288} (\bibinfo {year} {2003})},\ \Eprint {http://arxiv.org/abs/astro-ph/0212469} {arXiv:astro-ph/0212469 [astro-ph]} \BibitemShut {NoStop}%
\bibitem [{\citenamefont {{Quimby}}\ \emph {et~al.}(2011)\citenamefont {{Quimby}}, \citenamefont {{Kulkarni}}, \citenamefont {{Kasliwal}}, \citenamefont {{Gal-Yam}}, \citenamefont {{Arcavi}}, \citenamefont {{Sullivan}}, \citenamefont {{Nugent}}, \citenamefont {{Thomas}}, \citenamefont {{Howell}}, \citenamefont {{Nakar}}, \citenamefont {{Bildsten}}, \citenamefont {{Theissen}}, \citenamefont {{Law}}, \citenamefont {{Dekany}}, \citenamefont {{Rahmer}}, \citenamefont {{Hale}}, \citenamefont {{Smith}}, \citenamefont {{Ofek}}, \citenamefont {{Zolkower}}, \citenamefont {{Velur}}, \citenamefont {{Walters}}, \citenamefont {{Henning}}, \citenamefont {{Bui}}, \citenamefont {{McKenna}}, \citenamefont {{Poznanski}}, \citenamefont {{Cenko}},\ and\ \citenamefont {{Levitan}}}]{Quimby+2011_SLSN}%
  \BibitemOpen
  \bibfield  {author} {\bibinfo {author} {\bibfnamefont {R.~M.}\ \bibnamefont {{Quimby}}}, \bibinfo {author} {\bibfnamefont {S.~R.}\ \bibnamefont {{Kulkarni}}}, \bibinfo {author} {\bibfnamefont {M.~M.}\ \bibnamefont {{Kasliwal}}}, \bibinfo {author} {\bibfnamefont {A.}~\bibnamefont {{Gal-Yam}}}, \bibinfo {author} {\bibfnamefont {I.}~\bibnamefont {{Arcavi}}}, \bibinfo {author} {\bibfnamefont {M.}~\bibnamefont {{Sullivan}}}, \bibinfo {author} {\bibfnamefont {P.}~\bibnamefont {{Nugent}}}, \bibinfo {author} {\bibfnamefont {R.}~\bibnamefont {{Thomas}}}, \bibinfo {author} {\bibfnamefont {D.~A.}\ \bibnamefont {{Howell}}}, \bibinfo {author} {\bibfnamefont {E.}~\bibnamefont {{Nakar}}}, \bibinfo {author} {\bibfnamefont {L.}~\bibnamefont {{Bildsten}}}, \bibinfo {author} {\bibfnamefont {C.}~\bibnamefont {{Theissen}}}, \bibinfo {author} {\bibfnamefont {N.~M.}\ \bibnamefont {{Law}}}, \bibinfo {author} {\bibfnamefont {R.}~\bibnamefont {{Dekany}}}, \bibinfo {author} {\bibfnamefont {G.}~\bibnamefont {{Rahmer}}}, \bibinfo
  {author} {\bibfnamefont {D.}~\bibnamefont {{Hale}}}, \bibinfo {author} {\bibfnamefont {R.}~\bibnamefont {{Smith}}}, \bibinfo {author} {\bibfnamefont {E.~O.}\ \bibnamefont {{Ofek}}}, \bibinfo {author} {\bibfnamefont {J.}~\bibnamefont {{Zolkower}}}, \bibinfo {author} {\bibfnamefont {V.}~\bibnamefont {{Velur}}}, \bibinfo {author} {\bibfnamefont {R.}~\bibnamefont {{Walters}}}, \bibinfo {author} {\bibfnamefont {J.}~\bibnamefont {{Henning}}}, \bibinfo {author} {\bibfnamefont {K.}~\bibnamefont {{Bui}}}, \bibinfo {author} {\bibfnamefont {D.}~\bibnamefont {{McKenna}}}, \bibinfo {author} {\bibfnamefont {D.}~\bibnamefont {{Poznanski}}}, \bibinfo {author} {\bibfnamefont {S.~B.}\ \bibnamefont {{Cenko}}}, \ and\ \bibinfo {author} {\bibfnamefont {D.}~\bibnamefont {{Levitan}}},\ }\href {\doibase 10.1038/nature10095} {\bibfield  {journal} {\bibinfo  {journal} {\nat}\ }\textbf {\bibinfo {volume} {474}},\ \bibinfo {pages} {487} (\bibinfo {year} {2011})},\ \Eprint {http://arxiv.org/abs/0910.0059} {arXiv:0910.0059
  [astro-ph.CO]} \BibitemShut {NoStop}%
\bibitem [{\citenamefont {{Smith}}\ \emph {et~al.}(2012)\citenamefont {{Smith}}, \citenamefont {{Cenko}}, \citenamefont {{Butler}}, \citenamefont {{Bloom}}, \citenamefont {{Kasliwal}}, \citenamefont {{Horesh}}, \citenamefont {{Kulkarni}}, \citenamefont {{Law}}, \citenamefont {{Nugent}}, \citenamefont {{Ofek}}, \citenamefont {{Poznanski}}, \citenamefont {{Quimby}}, \citenamefont {{Sesar}}, \citenamefont {{Ben-Ami}}, \citenamefont {{Arcavi}}, \citenamefont {{Gal-Yam}}, \citenamefont {{Polishook}}, \citenamefont {{Xu}}, \citenamefont {{Yaron}}, \citenamefont {{Frail}},\ and\ \citenamefont {{Sullivan}}}]{Smith+2012_SN2010jp_PTF10aaxi_Halpha_triple}%
  \BibitemOpen
  \bibfield  {author} {\bibinfo {author} {\bibfnamefont {N.}~\bibnamefont {{Smith}}}, \bibinfo {author} {\bibfnamefont {S.~B.}\ \bibnamefont {{Cenko}}}, \bibinfo {author} {\bibfnamefont {N.}~\bibnamefont {{Butler}}}, \bibinfo {author} {\bibfnamefont {J.~S.}\ \bibnamefont {{Bloom}}}, \bibinfo {author} {\bibfnamefont {M.~M.}\ \bibnamefont {{Kasliwal}}}, \bibinfo {author} {\bibfnamefont {A.}~\bibnamefont {{Horesh}}}, \bibinfo {author} {\bibfnamefont {S.~R.}\ \bibnamefont {{Kulkarni}}}, \bibinfo {author} {\bibfnamefont {N.~M.}\ \bibnamefont {{Law}}}, \bibinfo {author} {\bibfnamefont {P.~E.}\ \bibnamefont {{Nugent}}}, \bibinfo {author} {\bibfnamefont {E.~O.}\ \bibnamefont {{Ofek}}}, \bibinfo {author} {\bibfnamefont {D.}~\bibnamefont {{Poznanski}}}, \bibinfo {author} {\bibfnamefont {R.~M.}\ \bibnamefont {{Quimby}}}, \bibinfo {author} {\bibfnamefont {B.}~\bibnamefont {{Sesar}}}, \bibinfo {author} {\bibfnamefont {S.}~\bibnamefont {{Ben-Ami}}}, \bibinfo {author} {\bibfnamefont {I.}~\bibnamefont {{Arcavi}}}, \bibinfo
  {author} {\bibfnamefont {A.}~\bibnamefont {{Gal-Yam}}}, \bibinfo {author} {\bibfnamefont {D.}~\bibnamefont {{Polishook}}}, \bibinfo {author} {\bibfnamefont {D.}~\bibnamefont {{Xu}}}, \bibinfo {author} {\bibfnamefont {O.}~\bibnamefont {{Yaron}}}, \bibinfo {author} {\bibfnamefont {D.~A.}\ \bibnamefont {{Frail}}}, \ and\ \bibinfo {author} {\bibfnamefont {M.}~\bibnamefont {{Sullivan}}},\ }\href {\doibase 10.1111/j.1365-2966.2011.20104.x} {\bibfield  {journal} {\bibinfo  {journal} {\mnras}\ }\textbf {\bibinfo {volume} {420}},\ \bibinfo {pages} {1135} (\bibinfo {year} {2012})},\ \Eprint {http://arxiv.org/abs/1108.2868} {arXiv:1108.2868 [astro-ph.HE]} \BibitemShut {NoStop}%
\bibitem [{\citenamefont {{Drout}}\ \emph {et~al.}(2014)\citenamefont {{Drout}}, \citenamefont {{Chornock}}, \citenamefont {{Soderberg}}, \citenamefont {{Sand ers}}, \citenamefont {{McKinnon}}, \citenamefont {{Rest}}, \citenamefont {{Foley}}, \citenamefont {{Milisavljevic}}, \citenamefont {{Margutti}},\ and\ \citenamefont {{Berger}}}]{Drout+2014_Rapidly_Evolving}%
  \BibitemOpen
  \bibfield  {author} {\bibinfo {author} {\bibfnamefont {M.~R.}\ \bibnamefont {{Drout}}}, \bibinfo {author} {\bibfnamefont {R.}~\bibnamefont {{Chornock}}}, \bibinfo {author} {\bibfnamefont {A.~M.}\ \bibnamefont {{Soderberg}}}, \bibinfo {author} {\bibfnamefont {N.~E.}\ \bibnamefont {{Sand ers}}}, \bibinfo {author} {\bibfnamefont {R.}~\bibnamefont {{McKinnon}}}, \bibinfo {author} {\bibfnamefont {A.}~\bibnamefont {{Rest}}}, \bibinfo {author} {\bibfnamefont {R.~J.}\ \bibnamefont {{Foley}}}, \bibinfo {author} {\bibfnamefont {D.}~\bibnamefont {{Milisavljevic}}}, \bibinfo {author} {\bibfnamefont {R.}~\bibnamefont {{Margutti}}}, \ and\ \bibinfo {author} {\bibfnamefont {E.}~\bibnamefont {{Berger}}},\ }\href {\doibase 10.1088/0004-637X/794/1/23} {\bibfield  {journal} {\bibinfo  {journal} {\apj}\ }\textbf {\bibinfo {volume} {794}},\ \bibinfo {eid} {23} (\bibinfo {year} {2014})},\ \Eprint {http://arxiv.org/abs/1405.3668} {arXiv:1405.3668 [astro-ph.HE]} \BibitemShut {NoStop}%
\bibitem [{\citenamefont {{Arcavi}}\ \emph {et~al.}(2016)\citenamefont {{Arcavi}}, \citenamefont {{Wolf}}, \citenamefont {{Howell}}, \citenamefont {{Bildsten}}, \citenamefont {{Leloudas}}, \citenamefont {{Hardin}}, \citenamefont {{Prajs}}, \citenamefont {{Perley}}, \citenamefont {{Svirski}}, \citenamefont {{Gal-Yam}}, \citenamefont {{Katz}}, \citenamefont {{McCully}}, \citenamefont {{Cenko}}, \citenamefont {{Lidman}}, \citenamefont {{Sullivan}}, \citenamefont {{Valenti}}, \citenamefont {{Astier}}, \citenamefont {{Balland}}, \citenamefont {{Carlberg}}, \citenamefont {{Conley}}, \citenamefont {{Fouchez}}, \citenamefont {{Guy}}, \citenamefont {{Pain}}, \citenamefont {{Palanque-Delabrouille}}, \citenamefont {{Perrett}}, \citenamefont {{Pritchet}}, \citenamefont {{Regnault}}, \citenamefont {{Rich}},\ and\ \citenamefont {{Ruhlmann-Kleider}}}]{Arcavi+2016_FastTransients}%
  \BibitemOpen
  \bibfield  {author} {\bibinfo {author} {\bibfnamefont {I.}~\bibnamefont {{Arcavi}}}, \bibinfo {author} {\bibfnamefont {W.~M.}\ \bibnamefont {{Wolf}}}, \bibinfo {author} {\bibfnamefont {D.~A.}\ \bibnamefont {{Howell}}}, \bibinfo {author} {\bibfnamefont {L.}~\bibnamefont {{Bildsten}}}, \bibinfo {author} {\bibfnamefont {G.}~\bibnamefont {{Leloudas}}}, \bibinfo {author} {\bibfnamefont {D.}~\bibnamefont {{Hardin}}}, \bibinfo {author} {\bibfnamefont {S.}~\bibnamefont {{Prajs}}}, \bibinfo {author} {\bibfnamefont {D.~A.}\ \bibnamefont {{Perley}}}, \bibinfo {author} {\bibfnamefont {G.}~\bibnamefont {{Svirski}}}, \bibinfo {author} {\bibfnamefont {A.}~\bibnamefont {{Gal-Yam}}}, \bibinfo {author} {\bibfnamefont {B.}~\bibnamefont {{Katz}}}, \bibinfo {author} {\bibfnamefont {C.}~\bibnamefont {{McCully}}}, \bibinfo {author} {\bibfnamefont {S.~B.}\ \bibnamefont {{Cenko}}}, \bibinfo {author} {\bibfnamefont {C.}~\bibnamefont {{Lidman}}}, \bibinfo {author} {\bibfnamefont {M.}~\bibnamefont {{Sullivan}}}, \bibinfo {author}
  {\bibfnamefont {S.}~\bibnamefont {{Valenti}}}, \bibinfo {author} {\bibfnamefont {P.}~\bibnamefont {{Astier}}}, \bibinfo {author} {\bibfnamefont {C.}~\bibnamefont {{Balland}}}, \bibinfo {author} {\bibfnamefont {R.~G.}\ \bibnamefont {{Carlberg}}}, \bibinfo {author} {\bibfnamefont {A.}~\bibnamefont {{Conley}}}, \bibinfo {author} {\bibfnamefont {D.}~\bibnamefont {{Fouchez}}}, \bibinfo {author} {\bibfnamefont {J.}~\bibnamefont {{Guy}}}, \bibinfo {author} {\bibfnamefont {R.}~\bibnamefont {{Pain}}}, \bibinfo {author} {\bibfnamefont {N.}~\bibnamefont {{Palanque-Delabrouille}}}, \bibinfo {author} {\bibfnamefont {K.}~\bibnamefont {{Perrett}}}, \bibinfo {author} {\bibfnamefont {C.~J.}\ \bibnamefont {{Pritchet}}}, \bibinfo {author} {\bibfnamefont {N.}~\bibnamefont {{Regnault}}}, \bibinfo {author} {\bibfnamefont {J.}~\bibnamefont {{Rich}}}, \ and\ \bibinfo {author} {\bibfnamefont {V.}~\bibnamefont {{Ruhlmann-Kleider}}},\ }\href {\doibase 10.3847/0004-637X/819/1/35} {\bibfield  {journal} {\bibinfo  {journal} {\apj}\
  }\textbf {\bibinfo {volume} {819}},\ \bibinfo {eid} {35} (\bibinfo {year} {2016})},\ \Eprint {http://arxiv.org/abs/1511.00704} {arXiv:1511.00704 [astro-ph.CO]} \BibitemShut {NoStop}%
\bibitem [{\citenamefont {{Thompson}}\ \emph {et~al.}(2004)\citenamefont {{Thompson}}, \citenamefont {{Chang}},\ and\ \citenamefont {{Quataert}}}]{Thompson+2004ApJ_SN_MagnetarSpinDown_GRB}%
  \BibitemOpen
  \bibfield  {author} {\bibinfo {author} {\bibfnamefont {T.~A.}\ \bibnamefont {{Thompson}}}, \bibinfo {author} {\bibfnamefont {P.}~\bibnamefont {{Chang}}}, \ and\ \bibinfo {author} {\bibfnamefont {E.}~\bibnamefont {{Quataert}}},\ }\href {\doibase 10.1086/421969} {\bibfield  {journal} {\bibinfo  {journal} {\apj}\ }\textbf {\bibinfo {volume} {611}},\ \bibinfo {pages} {380} (\bibinfo {year} {2004})},\ \Eprint {http://arxiv.org/abs/astro-ph/0401555} {arXiv:astro-ph/0401555 [astro-ph]} \BibitemShut {NoStop}%
\bibitem [{\citenamefont {{Woosley}}(2010)}]{Woosley2010ApJ_BrightSN_Magnetar}%
  \BibitemOpen
  \bibfield  {author} {\bibinfo {author} {\bibfnamefont {S.~E.}\ \bibnamefont {{Woosley}}},\ }\href {\doibase 10.1088/2041-8205/719/2/L204} {\bibfield  {journal} {\bibinfo  {journal} {\apjl}\ }\textbf {\bibinfo {volume} {719}},\ \bibinfo {pages} {L204} (\bibinfo {year} {2010})},\ \Eprint {http://arxiv.org/abs/0911.0698} {arXiv:0911.0698 [astro-ph.HE]} \BibitemShut {NoStop}%
\bibitem [{\citenamefont {{Kasen}}\ \emph {et~al.}(2016)\citenamefont {{Kasen}}, \citenamefont {{Metzger}},\ and\ \citenamefont {{Bildsten}}}]{Kasen+2016ApJ_MagnetarDrivenShockBreakout_SupernovaLightCurves}%
  \BibitemOpen
  \bibfield  {author} {\bibinfo {author} {\bibfnamefont {D.}~\bibnamefont {{Kasen}}}, \bibinfo {author} {\bibfnamefont {B.~D.}\ \bibnamefont {{Metzger}}}, \ and\ \bibinfo {author} {\bibfnamefont {L.}~\bibnamefont {{Bildsten}}},\ }\href {\doibase 10.3847/0004-637X/821/1/36} {\bibfield  {journal} {\bibinfo  {journal} {\apj}\ }\textbf {\bibinfo {volume} {821}},\ \bibinfo {eid} {36} (\bibinfo {year} {2016})},\ \Eprint {http://arxiv.org/abs/1507.03645} {arXiv:1507.03645 [astro-ph.HE]} \BibitemShut {NoStop}%
\bibitem [{\citenamefont {{Dexter}}\ and\ \citenamefont {{Kasen}}(2013)}]{Dexter+Kasen2013ApJ_SN_Powered_FallbackAcretion}%
  \BibitemOpen
  \bibfield  {author} {\bibinfo {author} {\bibfnamefont {J.}~\bibnamefont {{Dexter}}}\ and\ \bibinfo {author} {\bibfnamefont {D.}~\bibnamefont {{Kasen}}},\ }\href {\doibase 10.1088/0004-637X/772/1/30} {\bibfield  {journal} {\bibinfo  {journal} {\apj}\ }\textbf {\bibinfo {volume} {772}},\ \bibinfo {eid} {30} (\bibinfo {year} {2013})},\ \Eprint {http://arxiv.org/abs/1210.7240} {arXiv:1210.7240 [astro-ph.HE]} \BibitemShut {NoStop}%
\bibitem [{\citenamefont {{Chen}}\ \emph {et~al.}(2024)\citenamefont {{Chen}}, \citenamefont {{Gal-Yam}}, \citenamefont {{Sollerman}}, \citenamefont {{Schulze}}, \citenamefont {{Post}}, \citenamefont {{Liu}}, \citenamefont {{Ofek}}, \citenamefont {{Das}}, \citenamefont {{Fremling}}, \citenamefont {{Horesh}}, \citenamefont {{Katz}}, \citenamefont {{Kushnir}}, \citenamefont {{Kasliwal}}, \citenamefont {{Kulkarni}}, \citenamefont {{Liu}}, \citenamefont {{Liu}}, \citenamefont {{Miller}}, \citenamefont {{Rose}}, \citenamefont {{Waxman}}, \citenamefont {{Yang}}, \citenamefont {{Yao}}, \citenamefont {{Zackay}}, \citenamefont {{Bellm}}, \citenamefont {{Dekany}}, \citenamefont {{Drake}}, \citenamefont {{Fang}}, \citenamefont {{Fynbo}}, \citenamefont {{Groom}}, \citenamefont {{Helou}}, \citenamefont {{Irani}}, \citenamefont {{Jegou du Laz}}, \citenamefont {{Liu}}, \citenamefont {{Mazzali}}, \citenamefont {{Neill}}, \citenamefont {{Qin}}, \citenamefont {{Riddle}}, \citenamefont {{Sharon}}, \citenamefont {{Strotjohann}},
  \citenamefont {{Wold}},\ and\ \citenamefont {{Yan}}}]{Chen+2024Natur12days_Period_Supernova}%
  \BibitemOpen
  \bibfield  {author} {\bibinfo {author} {\bibfnamefont {P.}~\bibnamefont {{Chen}}}, \bibinfo {author} {\bibfnamefont {A.}~\bibnamefont {{Gal-Yam}}}, \bibinfo {author} {\bibfnamefont {J.}~\bibnamefont {{Sollerman}}}, \bibinfo {author} {\bibfnamefont {S.}~\bibnamefont {{Schulze}}}, \bibinfo {author} {\bibfnamefont {R.~S.}\ \bibnamefont {{Post}}}, \bibinfo {author} {\bibfnamefont {C.}~\bibnamefont {{Liu}}}, \bibinfo {author} {\bibfnamefont {E.~O.}\ \bibnamefont {{Ofek}}}, \bibinfo {author} {\bibfnamefont {K.~K.}\ \bibnamefont {{Das}}}, \bibinfo {author} {\bibfnamefont {C.}~\bibnamefont {{Fremling}}}, \bibinfo {author} {\bibfnamefont {A.}~\bibnamefont {{Horesh}}}, \bibinfo {author} {\bibfnamefont {B.}~\bibnamefont {{Katz}}}, \bibinfo {author} {\bibfnamefont {D.}~\bibnamefont {{Kushnir}}}, \bibinfo {author} {\bibfnamefont {M.~M.}\ \bibnamefont {{Kasliwal}}}, \bibinfo {author} {\bibfnamefont {S.~R.}\ \bibnamefont {{Kulkarni}}}, \bibinfo {author} {\bibfnamefont {D.}~\bibnamefont {{Liu}}}, \bibinfo {author}
  {\bibfnamefont {X.}~\bibnamefont {{Liu}}}, \bibinfo {author} {\bibfnamefont {A.~A.}\ \bibnamefont {{Miller}}}, \bibinfo {author} {\bibfnamefont {K.}~\bibnamefont {{Rose}}}, \bibinfo {author} {\bibfnamefont {E.}~\bibnamefont {{Waxman}}}, \bibinfo {author} {\bibfnamefont {S.}~\bibnamefont {{Yang}}}, \bibinfo {author} {\bibfnamefont {Y.}~\bibnamefont {{Yao}}}, \bibinfo {author} {\bibfnamefont {B.}~\bibnamefont {{Zackay}}}, \bibinfo {author} {\bibfnamefont {E.~C.}\ \bibnamefont {{Bellm}}}, \bibinfo {author} {\bibfnamefont {R.}~\bibnamefont {{Dekany}}}, \bibinfo {author} {\bibfnamefont {A.~J.}\ \bibnamefont {{Drake}}}, \bibinfo {author} {\bibfnamefont {Y.}~\bibnamefont {{Fang}}}, \bibinfo {author} {\bibfnamefont {J.~P.~U.}\ \bibnamefont {{Fynbo}}}, \bibinfo {author} {\bibfnamefont {S.~L.}\ \bibnamefont {{Groom}}}, \bibinfo {author} {\bibfnamefont {G.}~\bibnamefont {{Helou}}}, \bibinfo {author} {\bibfnamefont {I.}~\bibnamefont {{Irani}}}, \bibinfo {author} {\bibfnamefont {T.}~\bibnamefont {{Jegou du Laz}}},
  \bibinfo {author} {\bibfnamefont {X.}~\bibnamefont {{Liu}}}, \bibinfo {author} {\bibfnamefont {P.~A.}\ \bibnamefont {{Mazzali}}}, \bibinfo {author} {\bibfnamefont {J.~D.}\ \bibnamefont {{Neill}}}, \bibinfo {author} {\bibfnamefont {Y.-J.}\ \bibnamefont {{Qin}}}, \bibinfo {author} {\bibfnamefont {R.~L.}\ \bibnamefont {{Riddle}}}, \bibinfo {author} {\bibfnamefont {A.}~\bibnamefont {{Sharon}}}, \bibinfo {author} {\bibfnamefont {N.~L.}\ \bibnamefont {{Strotjohann}}}, \bibinfo {author} {\bibfnamefont {A.}~\bibnamefont {{Wold}}}, \ and\ \bibinfo {author} {\bibfnamefont {L.}~\bibnamefont {{Yan}}},\ }\href {\doibase 10.1038/s41586-023-06787-x} {\bibfield  {journal} {\bibinfo  {journal} {\nat}\ }\textbf {\bibinfo {volume} {625}},\ \bibinfo {pages} {253} (\bibinfo {year} {2024})},\ \Eprint {http://arxiv.org/abs/2310.07784} {arXiv:2310.07784 [astro-ph.HE]} \BibitemShut {NoStop}%
\bibitem [{\citenamefont {{Ho}}\ \emph {et~al.}(2023{\natexlab{a}})\citenamefont {{Ho}}, \citenamefont {{Perley}}, \citenamefont {{Chen}}, \citenamefont {{Schulze}}, \citenamefont {{Dhillon}}, \citenamefont {{Kumar}}, \citenamefont {{Suresh}}, \citenamefont {{Swain}}, \citenamefont {{Bremer}}, \citenamefont {{Smartt}}, \citenamefont {{Anderson}}, \citenamefont {{Anupama}}, \citenamefont {{Awiphan}}, \citenamefont {{Barway}}, \citenamefont {{Bellm}}, \citenamefont {{Ben-Ami}}, \citenamefont {{Bhalerao}}, \citenamefont {{de Boer}}, \citenamefont {{Brink}}, \citenamefont {{Burruss}}, \citenamefont {{Chandra}}, \citenamefont {{Chen}}, \citenamefont {{Chen}}, \citenamefont {{Cooke}}, \citenamefont {{Coughlin}}, \citenamefont {{Das}}, \citenamefont {{Drake}}, \citenamefont {{Filippenko}}, \citenamefont {{Freeburn}}, \citenamefont {{Fremling}}, \citenamefont {{Fulton}}, \citenamefont {{Gal-Yam}}, \citenamefont {{Galbany}}, \citenamefont {{Gao}}, \citenamefont {{Graham}}, \citenamefont {{Gromadzki}}, \citenamefont
  {{Guti{\'e}rrez}}, \citenamefont {{Hinds}}, \citenamefont {{Inserra}}, \citenamefont {{A J}}, \citenamefont {{Karambelkar}}, \citenamefont {{Kasliwal}}, \citenamefont {{Kulkarni}}, \citenamefont {{M{\"u}ller-Bravo}}, \citenamefont {{Magnier}}, \citenamefont {{Mahabal}}, \citenamefont {{Moore}}, \citenamefont {{Ngeow}}, \citenamefont {{Nicholl}}, \citenamefont {{Ofek}}, \citenamefont {{Omand}}, \citenamefont {{Onori}}, \citenamefont {{Pan}}, \citenamefont {{Pessi}}, \citenamefont {{Petitpas}}, \citenamefont {{Polishook}}, \citenamefont {{Poshyachinda}}, \citenamefont {{Pursiainen}}, \citenamefont {{Riddle}}, \citenamefont {{Rodriguez}}, \citenamefont {{Rusholme}}, \citenamefont {{Segre}}, \citenamefont {{Sharma}}, \citenamefont {{Smith}}, \citenamefont {{Sollerman}}, \citenamefont {{Srivastav}}, \citenamefont {{Strotjohann}}, \citenamefont {{Suhr}}, \citenamefont {{Svinkin}}, \citenamefont {{Wang}}, \citenamefont {{Wiseman}}, \citenamefont {{Wold}}, \citenamefont {{Yang}}, \citenamefont {{Yang}},
  \citenamefont {{Yao}}, \citenamefont {{Young}},\ and\ \citenamefont {{Zheng}}}]{Ho+2023Natur_MinuteTimeScaleFlares_Transient}%
  \BibitemOpen
  \bibfield  {author} {\bibinfo {author} {\bibfnamefont {A.~Y.~Q.}\ \bibnamefont {{Ho}}}, \bibinfo {author} {\bibfnamefont {D.~A.}\ \bibnamefont {{Perley}}}, \bibinfo {author} {\bibfnamefont {P.}~\bibnamefont {{Chen}}}, \bibinfo {author} {\bibfnamefont {S.}~\bibnamefont {{Schulze}}}, \bibinfo {author} {\bibfnamefont {V.}~\bibnamefont {{Dhillon}}}, \bibinfo {author} {\bibfnamefont {H.}~\bibnamefont {{Kumar}}}, \bibinfo {author} {\bibfnamefont {A.}~\bibnamefont {{Suresh}}}, \bibinfo {author} {\bibfnamefont {V.}~\bibnamefont {{Swain}}}, \bibinfo {author} {\bibfnamefont {M.}~\bibnamefont {{Bremer}}}, \bibinfo {author} {\bibfnamefont {S.~J.}\ \bibnamefont {{Smartt}}}, \bibinfo {author} {\bibfnamefont {J.~P.}\ \bibnamefont {{Anderson}}}, \bibinfo {author} {\bibfnamefont {G.~C.}\ \bibnamefont {{Anupama}}}, \bibinfo {author} {\bibfnamefont {S.}~\bibnamefont {{Awiphan}}}, \bibinfo {author} {\bibfnamefont {S.}~\bibnamefont {{Barway}}}, \bibinfo {author} {\bibfnamefont {E.~C.}\ \bibnamefont {{Bellm}}}, \bibinfo {author}
  {\bibfnamefont {S.}~\bibnamefont {{Ben-Ami}}}, \bibinfo {author} {\bibfnamefont {V.}~\bibnamefont {{Bhalerao}}}, \bibinfo {author} {\bibfnamefont {T.}~\bibnamefont {{de Boer}}}, \bibinfo {author} {\bibfnamefont {T.~G.}\ \bibnamefont {{Brink}}}, \bibinfo {author} {\bibfnamefont {R.}~\bibnamefont {{Burruss}}}, \bibinfo {author} {\bibfnamefont {P.}~\bibnamefont {{Chandra}}}, \bibinfo {author} {\bibfnamefont {T.-W.}\ \bibnamefont {{Chen}}}, \bibinfo {author} {\bibfnamefont {W.-P.}\ \bibnamefont {{Chen}}}, \bibinfo {author} {\bibfnamefont {J.}~\bibnamefont {{Cooke}}}, \bibinfo {author} {\bibfnamefont {M.~W.}\ \bibnamefont {{Coughlin}}}, \bibinfo {author} {\bibfnamefont {K.~K.}\ \bibnamefont {{Das}}}, \bibinfo {author} {\bibfnamefont {A.~J.}\ \bibnamefont {{Drake}}}, \bibinfo {author} {\bibfnamefont {A.~V.}\ \bibnamefont {{Filippenko}}}, \bibinfo {author} {\bibfnamefont {J.}~\bibnamefont {{Freeburn}}}, \bibinfo {author} {\bibfnamefont {C.}~\bibnamefont {{Fremling}}}, \bibinfo {author} {\bibfnamefont {M.~D.}\
  \bibnamefont {{Fulton}}}, \bibinfo {author} {\bibfnamefont {A.}~\bibnamefont {{Gal-Yam}}}, \bibinfo {author} {\bibfnamefont {L.}~\bibnamefont {{Galbany}}}, \bibinfo {author} {\bibfnamefont {H.}~\bibnamefont {{Gao}}}, \bibinfo {author} {\bibfnamefont {M.~J.}\ \bibnamefont {{Graham}}}, \bibinfo {author} {\bibfnamefont {M.}~\bibnamefont {{Gromadzki}}}, \bibinfo {author} {\bibfnamefont {C.~P.}\ \bibnamefont {{Guti{\'e}rrez}}}, \bibinfo {author} {\bibfnamefont {K.~R.}\ \bibnamefont {{Hinds}}}, \bibinfo {author} {\bibfnamefont {C.}~\bibnamefont {{Inserra}}}, \bibinfo {author} {\bibfnamefont {N.}~\bibnamefont {{A J}}}, \bibinfo {author} {\bibfnamefont {V.}~\bibnamefont {{Karambelkar}}}, \bibinfo {author} {\bibfnamefont {M.~M.}\ \bibnamefont {{Kasliwal}}}, \bibinfo {author} {\bibfnamefont {S.}~\bibnamefont {{Kulkarni}}}, \bibinfo {author} {\bibfnamefont {T.~E.}\ \bibnamefont {{M{\"u}ller-Bravo}}}, \bibinfo {author} {\bibfnamefont {E.~A.}\ \bibnamefont {{Magnier}}}, \bibinfo {author} {\bibfnamefont {A.~A.}\
  \bibnamefont {{Mahabal}}}, \bibinfo {author} {\bibfnamefont {T.}~\bibnamefont {{Moore}}}, \bibinfo {author} {\bibfnamefont {C.-C.}\ \bibnamefont {{Ngeow}}}, \bibinfo {author} {\bibfnamefont {M.}~\bibnamefont {{Nicholl}}}, \bibinfo {author} {\bibfnamefont {E.~O.}\ \bibnamefont {{Ofek}}}, \bibinfo {author} {\bibfnamefont {C.~M.~B.}\ \bibnamefont {{Omand}}}, \bibinfo {author} {\bibfnamefont {F.}~\bibnamefont {{Onori}}}, \bibinfo {author} {\bibfnamefont {Y.-C.}\ \bibnamefont {{Pan}}}, \bibinfo {author} {\bibfnamefont {P.~J.}\ \bibnamefont {{Pessi}}}, \bibinfo {author} {\bibfnamefont {G.}~\bibnamefont {{Petitpas}}}, \bibinfo {author} {\bibfnamefont {D.}~\bibnamefont {{Polishook}}}, \bibinfo {author} {\bibfnamefont {S.}~\bibnamefont {{Poshyachinda}}}, \bibinfo {author} {\bibfnamefont {M.}~\bibnamefont {{Pursiainen}}}, \bibinfo {author} {\bibfnamefont {R.}~\bibnamefont {{Riddle}}}, \bibinfo {author} {\bibfnamefont {A.~C.}\ \bibnamefont {{Rodriguez}}}, \bibinfo {author} {\bibfnamefont {B.}~\bibnamefont
  {{Rusholme}}}, \bibinfo {author} {\bibfnamefont {E.}~\bibnamefont {{Segre}}}, \bibinfo {author} {\bibfnamefont {Y.}~\bibnamefont {{Sharma}}}, \bibinfo {author} {\bibfnamefont {K.~W.}\ \bibnamefont {{Smith}}}, \bibinfo {author} {\bibfnamefont {J.}~\bibnamefont {{Sollerman}}}, \bibinfo {author} {\bibfnamefont {S.}~\bibnamefont {{Srivastav}}}, \bibinfo {author} {\bibfnamefont {N.~L.}\ \bibnamefont {{Strotjohann}}}, \bibinfo {author} {\bibfnamefont {M.}~\bibnamefont {{Suhr}}}, \bibinfo {author} {\bibfnamefont {D.}~\bibnamefont {{Svinkin}}}, \bibinfo {author} {\bibfnamefont {Y.}~\bibnamefont {{Wang}}}, \bibinfo {author} {\bibfnamefont {P.}~\bibnamefont {{Wiseman}}}, \bibinfo {author} {\bibfnamefont {A.}~\bibnamefont {{Wold}}}, \bibinfo {author} {\bibfnamefont {S.}~\bibnamefont {{Yang}}}, \bibinfo {author} {\bibfnamefont {Y.}~\bibnamefont {{Yang}}}, \bibinfo {author} {\bibfnamefont {Y.}~\bibnamefont {{Yao}}}, \bibinfo {author} {\bibfnamefont {D.~R.}\ \bibnamefont {{Young}}}, \ and\ \bibinfo {author}
  {\bibfnamefont {W.}~\bibnamefont {{Zheng}}},\ }\href {\doibase 10.1038/s41586-023-06673-6} {\bibfield  {journal} {\bibinfo  {journal} {\nat}\ }\textbf {\bibinfo {volume} {623}},\ \bibinfo {pages} {927} (\bibinfo {year} {2023}{\natexlab{a}})},\ \Eprint {http://arxiv.org/abs/2311.10195} {arXiv:2311.10195 [astro-ph.HE]} \BibitemShut {NoStop}%
\bibitem [{\citenamefont {{Prentice}}\ \emph {et~al.}(2018)\citenamefont {{Prentice}}, \citenamefont {{Maguire}}, \citenamefont {{Smartt}}, \citenamefont {{Magee}}, \citenamefont {{Schady}}, \citenamefont {{Sim}}, \citenamefont {{Chen}}, \citenamefont {{Clark}}, \citenamefont {{Colin}}, \citenamefont {{Fulton}}, \citenamefont {{McBrien}}, \citenamefont {{O'Neill}}, \citenamefont {{Smith}}, \citenamefont {{Ashall}}, \citenamefont {{Chambers}}, \citenamefont {{Denneau}}, \citenamefont {{Flewelling}}, \citenamefont {{Heinze}}, \citenamefont {{Holoien}}, \citenamefont {{Huber}}, \citenamefont {{Kochanek}}, \citenamefont {{Mazzali}}, \citenamefont {{Prieto}}, \citenamefont {{Rest}}, \citenamefont {{Shappee}}, \citenamefont {{Stalder}}, \citenamefont {{Stanek}}, \citenamefont {{Stritzinger}}, \citenamefont {{Thompson}},\ and\ \citenamefont {{Tonry}}}]{Prentice+2018ApJ2018cow_Discovery}%
  \BibitemOpen
  \bibfield  {author} {\bibinfo {author} {\bibfnamefont {S.~J.}\ \bibnamefont {{Prentice}}}, \bibinfo {author} {\bibfnamefont {K.}~\bibnamefont {{Maguire}}}, \bibinfo {author} {\bibfnamefont {S.~J.}\ \bibnamefont {{Smartt}}}, \bibinfo {author} {\bibfnamefont {M.~R.}\ \bibnamefont {{Magee}}}, \bibinfo {author} {\bibfnamefont {P.}~\bibnamefont {{Schady}}}, \bibinfo {author} {\bibfnamefont {S.}~\bibnamefont {{Sim}}}, \bibinfo {author} {\bibfnamefont {T.~W.}\ \bibnamefont {{Chen}}}, \bibinfo {author} {\bibfnamefont {P.}~\bibnamefont {{Clark}}}, \bibinfo {author} {\bibfnamefont {C.}~\bibnamefont {{Colin}}}, \bibinfo {author} {\bibfnamefont {M.}~\bibnamefont {{Fulton}}}, \bibinfo {author} {\bibfnamefont {O.}~\bibnamefont {{McBrien}}}, \bibinfo {author} {\bibfnamefont {D.}~\bibnamefont {{O'Neill}}}, \bibinfo {author} {\bibfnamefont {K.~W.}\ \bibnamefont {{Smith}}}, \bibinfo {author} {\bibfnamefont {C.}~\bibnamefont {{Ashall}}}, \bibinfo {author} {\bibfnamefont {K.~C.}\ \bibnamefont {{Chambers}}}, \bibinfo {author}
  {\bibfnamefont {L.}~\bibnamefont {{Denneau}}}, \bibinfo {author} {\bibfnamefont {H.~A.}\ \bibnamefont {{Flewelling}}}, \bibinfo {author} {\bibfnamefont {A.}~\bibnamefont {{Heinze}}}, \bibinfo {author} {\bibfnamefont {T.~W.~S.}\ \bibnamefont {{Holoien}}}, \bibinfo {author} {\bibfnamefont {M.~E.}\ \bibnamefont {{Huber}}}, \bibinfo {author} {\bibfnamefont {C.~S.}\ \bibnamefont {{Kochanek}}}, \bibinfo {author} {\bibfnamefont {P.~A.}\ \bibnamefont {{Mazzali}}}, \bibinfo {author} {\bibfnamefont {J.~L.}\ \bibnamefont {{Prieto}}}, \bibinfo {author} {\bibfnamefont {A.}~\bibnamefont {{Rest}}}, \bibinfo {author} {\bibfnamefont {B.~J.}\ \bibnamefont {{Shappee}}}, \bibinfo {author} {\bibfnamefont {B.}~\bibnamefont {{Stalder}}}, \bibinfo {author} {\bibfnamefont {K.~Z.}\ \bibnamefont {{Stanek}}}, \bibinfo {author} {\bibfnamefont {M.~D.}\ \bibnamefont {{Stritzinger}}}, \bibinfo {author} {\bibfnamefont {T.~A.}\ \bibnamefont {{Thompson}}}, \ and\ \bibinfo {author} {\bibfnamefont {J.~L.}\ \bibnamefont {{Tonry}}},\ }\href
  {\doibase 10.3847/2041-8213/aadd90} {\bibfield  {journal} {\bibinfo  {journal} {\apjl}\ }\textbf {\bibinfo {volume} {865}},\ \bibinfo {eid} {L3} (\bibinfo {year} {2018})},\ \Eprint {http://arxiv.org/abs/1807.05965} {arXiv:1807.05965 [astro-ph.HE]} \BibitemShut {NoStop}%
\bibitem [{\citenamefont {{Ho}}\ \emph {et~al.}(2019)\citenamefont {{Ho}}, \citenamefont {{Phinney}}, \citenamefont {{Ravi}}, \citenamefont {{Kulkarni}}, \citenamefont {{Petitpas}}, \citenamefont {{Emonts}}, \citenamefont {{Bhalerao}}, \citenamefont {{Blundell}}, \citenamefont {{Cenko}}, \citenamefont {{Dobie}}, \citenamefont {{Howie}}, \citenamefont {{Kamraj}}, \citenamefont {{Kasliwal}}, \citenamefont {{Murphy}}, \citenamefont {{Perley}}, \citenamefont {{Sridharan}},\ and\ \citenamefont {{Yoon}}}]{Ho+2019_AT2018cow_Radio}%
  \BibitemOpen
  \bibfield  {author} {\bibinfo {author} {\bibfnamefont {A.~Y.~Q.}\ \bibnamefont {{Ho}}}, \bibinfo {author} {\bibfnamefont {E.~S.}\ \bibnamefont {{Phinney}}}, \bibinfo {author} {\bibfnamefont {V.}~\bibnamefont {{Ravi}}}, \bibinfo {author} {\bibfnamefont {S.~R.}\ \bibnamefont {{Kulkarni}}}, \bibinfo {author} {\bibfnamefont {G.}~\bibnamefont {{Petitpas}}}, \bibinfo {author} {\bibfnamefont {B.}~\bibnamefont {{Emonts}}}, \bibinfo {author} {\bibfnamefont {V.}~\bibnamefont {{Bhalerao}}}, \bibinfo {author} {\bibfnamefont {R.}~\bibnamefont {{Blundell}}}, \bibinfo {author} {\bibfnamefont {S.~B.}\ \bibnamefont {{Cenko}}}, \bibinfo {author} {\bibfnamefont {D.}~\bibnamefont {{Dobie}}}, \bibinfo {author} {\bibfnamefont {R.}~\bibnamefont {{Howie}}}, \bibinfo {author} {\bibfnamefont {N.}~\bibnamefont {{Kamraj}}}, \bibinfo {author} {\bibfnamefont {M.~M.}\ \bibnamefont {{Kasliwal}}}, \bibinfo {author} {\bibfnamefont {T.}~\bibnamefont {{Murphy}}}, \bibinfo {author} {\bibfnamefont {D.~A.}\ \bibnamefont {{Perley}}}, \bibinfo
  {author} {\bibfnamefont {T.~K.}\ \bibnamefont {{Sridharan}}}, \ and\ \bibinfo {author} {\bibfnamefont {I.}~\bibnamefont {{Yoon}}},\ }\href {\doibase 10.3847/1538-4357/aaf473} {\bibfield  {journal} {\bibinfo  {journal} {\apj}\ }\textbf {\bibinfo {volume} {871}},\ \bibinfo {eid} {73} (\bibinfo {year} {2019})},\ \Eprint {http://arxiv.org/abs/1810.10880} {arXiv:1810.10880 [astro-ph.HE]} \BibitemShut {NoStop}%
\bibitem [{\citenamefont {{Perley}}\ \emph {et~al.}(2019)\citenamefont {{Perley}}, \citenamefont {{Mazzali}}, \citenamefont {{Yan}}, \citenamefont {{Cenko}}, \citenamefont {{Gezari}}, \citenamefont {{Taggart}}, \citenamefont {{Blagorodnova}}, \citenamefont {{Fremling}}, \citenamefont {{Mockler}},\ and\ \citenamefont {{Singh}}}]{Perley+2019_SN2018cow}%
  \BibitemOpen
  \bibfield  {author} {\bibinfo {author} {\bibfnamefont {D.~A.}\ \bibnamefont {{Perley}}}, \bibinfo {author} {\bibfnamefont {P.~A.}\ \bibnamefont {{Mazzali}}}, \bibinfo {author} {\bibfnamefont {L.}~\bibnamefont {{Yan}}}, \bibinfo {author} {\bibfnamefont {S.~B.}\ \bibnamefont {{Cenko}}}, \bibinfo {author} {\bibfnamefont {S.}~\bibnamefont {{Gezari}}}, \bibinfo {author} {\bibfnamefont {K.}~\bibnamefont {{Taggart}}}, \bibinfo {author} {\bibfnamefont {N.}~\bibnamefont {{Blagorodnova}}}, \bibinfo {author} {\bibfnamefont {C.}~\bibnamefont {{Fremling}}}, \bibinfo {author} {\bibfnamefont {B.}~\bibnamefont {{Mockler}}}, \ and\ \bibinfo {author} {\bibfnamefont {A.}~\bibnamefont {{Singh}}},\ }\href {\doibase 10.1093/mnras/sty3420} {\bibfield  {journal} {\bibinfo  {journal} {\mnras}\ }\textbf {\bibinfo {volume} {484}},\ \bibinfo {pages} {1031} (\bibinfo {year} {2019})},\ \Eprint {http://arxiv.org/abs/1808.00969} {arXiv:1808.00969 [astro-ph.HE]} \BibitemShut {NoStop}%
\bibitem [{\citenamefont {{Margutti}}\ \emph {et~al.}(2019)\citenamefont {{Margutti}}, \citenamefont {{Metzger}}, \citenamefont {{Chornock}}, \citenamefont {{Vurm}}, \citenamefont {{Roth}}, \citenamefont {{Grefenstette}}, \citenamefont {{Savchenko}}, \citenamefont {{Cartier}}, \citenamefont {{Steiner}}, \citenamefont {{Terreran}}, \citenamefont {{Margalit}}, \citenamefont {{Migliori}}, \citenamefont {{Milisavljevic}}, \citenamefont {{Alexander}}, \citenamefont {{Bietenholz}}, \citenamefont {{Blanchard}}, \citenamefont {{Bozzo}}, \citenamefont {{Brethauer}}, \citenamefont {{Chilingarian}}, \citenamefont {{Coppejans}}, \citenamefont {{Ducci}}, \citenamefont {{Ferrigno}}, \citenamefont {{Fong}}, \citenamefont {{G{\"o}tz}}, \citenamefont {{Guidorzi}}, \citenamefont {{Hajela}}, \citenamefont {{Hurley}}, \citenamefont {{Kuulkers}}, \citenamefont {{Laurent}}, \citenamefont {{Mereghetti}}, \citenamefont {{Nicholl}}, \citenamefont {{Patnaude}}, \citenamefont {{Ubertini}}, \citenamefont {{Banovetz}}, \citenamefont
  {{Bartel}}, \citenamefont {{Berger}}, \citenamefont {{Coughlin}}, \citenamefont {{Eftekhari}}, \citenamefont {{Frederiks}}, \citenamefont {{Kozlova}}, \citenamefont {{Laskar}}, \citenamefont {{Svinkin}}, \citenamefont {{Drout}}, \citenamefont {{MacFadyen}},\ and\ \citenamefont {{Paterson}}}]{Margutti+2019ApJ_AT2018cow_Xray}%
  \BibitemOpen
  \bibfield  {author} {\bibinfo {author} {\bibfnamefont {R.}~\bibnamefont {{Margutti}}}, \bibinfo {author} {\bibfnamefont {B.~D.}\ \bibnamefont {{Metzger}}}, \bibinfo {author} {\bibfnamefont {R.}~\bibnamefont {{Chornock}}}, \bibinfo {author} {\bibfnamefont {I.}~\bibnamefont {{Vurm}}}, \bibinfo {author} {\bibfnamefont {N.}~\bibnamefont {{Roth}}}, \bibinfo {author} {\bibfnamefont {B.~W.}\ \bibnamefont {{Grefenstette}}}, \bibinfo {author} {\bibfnamefont {V.}~\bibnamefont {{Savchenko}}}, \bibinfo {author} {\bibfnamefont {R.}~\bibnamefont {{Cartier}}}, \bibinfo {author} {\bibfnamefont {J.~F.}\ \bibnamefont {{Steiner}}}, \bibinfo {author} {\bibfnamefont {G.}~\bibnamefont {{Terreran}}}, \bibinfo {author} {\bibfnamefont {B.}~\bibnamefont {{Margalit}}}, \bibinfo {author} {\bibfnamefont {G.}~\bibnamefont {{Migliori}}}, \bibinfo {author} {\bibfnamefont {D.}~\bibnamefont {{Milisavljevic}}}, \bibinfo {author} {\bibfnamefont {K.~D.}\ \bibnamefont {{Alexander}}}, \bibinfo {author} {\bibfnamefont {M.}~\bibnamefont
  {{Bietenholz}}}, \bibinfo {author} {\bibfnamefont {P.~K.}\ \bibnamefont {{Blanchard}}}, \bibinfo {author} {\bibfnamefont {E.}~\bibnamefont {{Bozzo}}}, \bibinfo {author} {\bibfnamefont {D.}~\bibnamefont {{Brethauer}}}, \bibinfo {author} {\bibfnamefont {I.~V.}\ \bibnamefont {{Chilingarian}}}, \bibinfo {author} {\bibfnamefont {D.~L.}\ \bibnamefont {{Coppejans}}}, \bibinfo {author} {\bibfnamefont {L.}~\bibnamefont {{Ducci}}}, \bibinfo {author} {\bibfnamefont {C.}~\bibnamefont {{Ferrigno}}}, \bibinfo {author} {\bibfnamefont {W.}~\bibnamefont {{Fong}}}, \bibinfo {author} {\bibfnamefont {D.}~\bibnamefont {{G{\"o}tz}}}, \bibinfo {author} {\bibfnamefont {C.}~\bibnamefont {{Guidorzi}}}, \bibinfo {author} {\bibfnamefont {A.}~\bibnamefont {{Hajela}}}, \bibinfo {author} {\bibfnamefont {K.}~\bibnamefont {{Hurley}}}, \bibinfo {author} {\bibfnamefont {E.}~\bibnamefont {{Kuulkers}}}, \bibinfo {author} {\bibfnamefont {P.}~\bibnamefont {{Laurent}}}, \bibinfo {author} {\bibfnamefont {S.}~\bibnamefont {{Mereghetti}}}, \bibinfo
  {author} {\bibfnamefont {M.}~\bibnamefont {{Nicholl}}}, \bibinfo {author} {\bibfnamefont {D.}~\bibnamefont {{Patnaude}}}, \bibinfo {author} {\bibfnamefont {P.}~\bibnamefont {{Ubertini}}}, \bibinfo {author} {\bibfnamefont {J.}~\bibnamefont {{Banovetz}}}, \bibinfo {author} {\bibfnamefont {N.}~\bibnamefont {{Bartel}}}, \bibinfo {author} {\bibfnamefont {E.}~\bibnamefont {{Berger}}}, \bibinfo {author} {\bibfnamefont {E.~R.}\ \bibnamefont {{Coughlin}}}, \bibinfo {author} {\bibfnamefont {T.}~\bibnamefont {{Eftekhari}}}, \bibinfo {author} {\bibfnamefont {D.~D.}\ \bibnamefont {{Frederiks}}}, \bibinfo {author} {\bibfnamefont {A.~V.}\ \bibnamefont {{Kozlova}}}, \bibinfo {author} {\bibfnamefont {T.}~\bibnamefont {{Laskar}}}, \bibinfo {author} {\bibfnamefont {D.~S.}\ \bibnamefont {{Svinkin}}}, \bibinfo {author} {\bibfnamefont {M.~R.}\ \bibnamefont {{Drout}}}, \bibinfo {author} {\bibfnamefont {A.}~\bibnamefont {{MacFadyen}}}, \ and\ \bibinfo {author} {\bibfnamefont {K.}~\bibnamefont {{Paterson}}},\ }\href {\doibase
  10.3847/1538-4357/aafa01} {\bibfield  {journal} {\bibinfo  {journal} {\apj}\ }\textbf {\bibinfo {volume} {872}},\ \bibinfo {eid} {18} (\bibinfo {year} {2019})},\ \Eprint {http://arxiv.org/abs/1810.10720} {arXiv:1810.10720 [astro-ph.HE]} \BibitemShut {NoStop}%
\bibitem [{\citenamefont {{Ho}}\ \emph {et~al.}(2023{\natexlab{b}})\citenamefont {{Ho}}, \citenamefont {{Perley}}, \citenamefont {{Gal-Yam}}, \citenamefont {{Lunnan}}, \citenamefont {{Sollerman}}, \citenamefont {{Schulze}}, \citenamefont {{Das}}, \citenamefont {{Dobie}}, \citenamefont {{Yao}}, \citenamefont {{Fremling}}, \citenamefont {{Adams}}, \citenamefont {{Anand}}, \citenamefont {{Andreoni}}, \citenamefont {{Bellm}}, \citenamefont {{Bruch}}, \citenamefont {{Burdge}}, \citenamefont {{Castro-Tirado}}, \citenamefont {{Dahiwale}}, \citenamefont {{De}}, \citenamefont {{Dekany}}, \citenamefont {{Drake}}, \citenamefont {{Duev}}, \citenamefont {{Graham}}, \citenamefont {{Helou}}, \citenamefont {{Kaplan}}, \citenamefont {{Karambelkar}}, \citenamefont {{Kasliwal}}, \citenamefont {{Kool}}, \citenamefont {{Kulkarni}}, \citenamefont {{Mahabal}}, \citenamefont {{Medford}}, \citenamefont {{Miller}}, \citenamefont {{Nordin}}, \citenamefont {{Ofek}}, \citenamefont {{Petitpas}}, \citenamefont {{Riddle}}, \citenamefont
  {{Sharma}}, \citenamefont {{Smith}}, \citenamefont {{Stewart}}, \citenamefont {{Taggart}}, \citenamefont {{Tartaglia}}, \citenamefont {{Tzanidakis}},\ and\ \citenamefont {{Winters}}}]{Ho+2023ApJ_SearchFBOT_ZTF_AT2018cow_Rate}%
  \BibitemOpen
  \bibfield  {author} {\bibinfo {author} {\bibfnamefont {A.~Y.~Q.}\ \bibnamefont {{Ho}}}, \bibinfo {author} {\bibfnamefont {D.~A.}\ \bibnamefont {{Perley}}}, \bibinfo {author} {\bibfnamefont {A.}~\bibnamefont {{Gal-Yam}}}, \bibinfo {author} {\bibfnamefont {R.}~\bibnamefont {{Lunnan}}}, \bibinfo {author} {\bibfnamefont {J.}~\bibnamefont {{Sollerman}}}, \bibinfo {author} {\bibfnamefont {S.}~\bibnamefont {{Schulze}}}, \bibinfo {author} {\bibfnamefont {K.~K.}\ \bibnamefont {{Das}}}, \bibinfo {author} {\bibfnamefont {D.}~\bibnamefont {{Dobie}}}, \bibinfo {author} {\bibfnamefont {Y.}~\bibnamefont {{Yao}}}, \bibinfo {author} {\bibfnamefont {C.}~\bibnamefont {{Fremling}}}, \bibinfo {author} {\bibfnamefont {S.}~\bibnamefont {{Adams}}}, \bibinfo {author} {\bibfnamefont {S.}~\bibnamefont {{Anand}}}, \bibinfo {author} {\bibfnamefont {I.}~\bibnamefont {{Andreoni}}}, \bibinfo {author} {\bibfnamefont {E.~C.}\ \bibnamefont {{Bellm}}}, \bibinfo {author} {\bibfnamefont {R.~J.}\ \bibnamefont {{Bruch}}}, \bibinfo {author}
  {\bibfnamefont {K.~B.}\ \bibnamefont {{Burdge}}}, \bibinfo {author} {\bibfnamefont {A.~J.}\ \bibnamefont {{Castro-Tirado}}}, \bibinfo {author} {\bibfnamefont {A.}~\bibnamefont {{Dahiwale}}}, \bibinfo {author} {\bibfnamefont {K.}~\bibnamefont {{De}}}, \bibinfo {author} {\bibfnamefont {R.}~\bibnamefont {{Dekany}}}, \bibinfo {author} {\bibfnamefont {A.~J.}\ \bibnamefont {{Drake}}}, \bibinfo {author} {\bibfnamefont {D.~A.}\ \bibnamefont {{Duev}}}, \bibinfo {author} {\bibfnamefont {M.~J.}\ \bibnamefont {{Graham}}}, \bibinfo {author} {\bibfnamefont {G.}~\bibnamefont {{Helou}}}, \bibinfo {author} {\bibfnamefont {D.~L.}\ \bibnamefont {{Kaplan}}}, \bibinfo {author} {\bibfnamefont {V.}~\bibnamefont {{Karambelkar}}}, \bibinfo {author} {\bibfnamefont {M.~M.}\ \bibnamefont {{Kasliwal}}}, \bibinfo {author} {\bibfnamefont {E.~C.}\ \bibnamefont {{Kool}}}, \bibinfo {author} {\bibfnamefont {S.~R.}\ \bibnamefont {{Kulkarni}}}, \bibinfo {author} {\bibfnamefont {A.~A.}\ \bibnamefont {{Mahabal}}}, \bibinfo {author}
  {\bibfnamefont {M.~S.}\ \bibnamefont {{Medford}}}, \bibinfo {author} {\bibfnamefont {A.~A.}\ \bibnamefont {{Miller}}}, \bibinfo {author} {\bibfnamefont {J.}~\bibnamefont {{Nordin}}}, \bibinfo {author} {\bibfnamefont {E.}~\bibnamefont {{Ofek}}}, \bibinfo {author} {\bibfnamefont {G.}~\bibnamefont {{Petitpas}}}, \bibinfo {author} {\bibfnamefont {R.}~\bibnamefont {{Riddle}}}, \bibinfo {author} {\bibfnamefont {Y.}~\bibnamefont {{Sharma}}}, \bibinfo {author} {\bibfnamefont {R.}~\bibnamefont {{Smith}}}, \bibinfo {author} {\bibfnamefont {A.~J.}\ \bibnamefont {{Stewart}}}, \bibinfo {author} {\bibfnamefont {K.}~\bibnamefont {{Taggart}}}, \bibinfo {author} {\bibfnamefont {L.}~\bibnamefont {{Tartaglia}}}, \bibinfo {author} {\bibfnamefont {A.}~\bibnamefont {{Tzanidakis}}}, \ and\ \bibinfo {author} {\bibfnamefont {J.~M.}\ \bibnamefont {{Winters}}},\ }\href {\doibase 10.3847/1538-4357/acc533} {\bibfield  {journal} {\bibinfo  {journal} {\apj}\ }\textbf {\bibinfo {volume} {949}},\ \bibinfo {eid} {120} (\bibinfo {year}
  {2023}{\natexlab{b}})},\ \Eprint {http://arxiv.org/abs/2105.08811} {arXiv:2105.08811 [astro-ph.HE]} \BibitemShut {NoStop}%
\bibitem [{\citenamefont {{Chen}}\ \emph {et~al.}(2023)\citenamefont {{Chen}}, \citenamefont {{Drout}}, \citenamefont {{Piro}}, \citenamefont {{Kilpatrick}}, \citenamefont {{Foley}}, \citenamefont {{Rojas-Bravo}},\ and\ \citenamefont {{Magee}}}]{Chen+2023ApJ_AT2018cow_HST_LatTimeUV}%
  \BibitemOpen
  \bibfield  {author} {\bibinfo {author} {\bibfnamefont {Y.}~\bibnamefont {{Chen}}}, \bibinfo {author} {\bibfnamefont {M.~R.}\ \bibnamefont {{Drout}}}, \bibinfo {author} {\bibfnamefont {A.~L.}\ \bibnamefont {{Piro}}}, \bibinfo {author} {\bibfnamefont {C.~D.}\ \bibnamefont {{Kilpatrick}}}, \bibinfo {author} {\bibfnamefont {R.~J.}\ \bibnamefont {{Foley}}}, \bibinfo {author} {\bibfnamefont {C.}~\bibnamefont {{Rojas-Bravo}}}, \ and\ \bibinfo {author} {\bibfnamefont {M.~R.}\ \bibnamefont {{Magee}}},\ }\href {\doibase 10.3847/1538-4357/ace964} {\bibfield  {journal} {\bibinfo  {journal} {\apj}\ }\textbf {\bibinfo {volume} {955}},\ \bibinfo {eid} {43} (\bibinfo {year} {2023})},\ \Eprint {http://arxiv.org/abs/2303.03501} {arXiv:2303.03501 [astro-ph.HE]} \BibitemShut {NoStop}%
\bibitem [{\citenamefont {{Perley}}\ \emph {et~al.}(2021)\citenamefont {{Perley}}, \citenamefont {{Ho}}, \citenamefont {{Yao}}, \citenamefont {{Fremling}}, \citenamefont {{Anderson}}, \citenamefont {{Schulze}}, \citenamefont {{Kumar}}, \citenamefont {{Anupama}}, \citenamefont {{Barway}}, \citenamefont {{Bellm}}, \citenamefont {{Bhalerao}}, \citenamefont {{Chen}}, \citenamefont {{Duev}}, \citenamefont {{Galbany}}, \citenamefont {{Graham}}, \citenamefont {{Gromadzki}}, \citenamefont {{Guti{\'e}rrez}}, \citenamefont {{Ihanec}}, \citenamefont {{Inserram}}, \citenamefont {{Kasliwal}}, \citenamefont {{Kool}}, \citenamefont {{Kulkarni}}, \citenamefont {{Laher}}, \citenamefont {{Masci}}, \citenamefont {{Neill}}, \citenamefont {{Nicholl}}, \citenamefont {{Pursiainen}}, \citenamefont {{van Roestel}}, \citenamefont {{Sharma}}, \citenamefont {{Sollerman}}, \citenamefont {{Walters}},\ and\ \citenamefont {{Wiseman}}}]{Perley+2020_AT2020xnd_FastLuminousTransient}%
  \BibitemOpen
  \bibfield  {author} {\bibinfo {author} {\bibfnamefont {D.~A.}\ \bibnamefont {{Perley}}}, \bibinfo {author} {\bibfnamefont {A.~Y.~Q.}\ \bibnamefont {{Ho}}}, \bibinfo {author} {\bibfnamefont {Y.}~\bibnamefont {{Yao}}}, \bibinfo {author} {\bibfnamefont {C.}~\bibnamefont {{Fremling}}}, \bibinfo {author} {\bibfnamefont {J.~P.}\ \bibnamefont {{Anderson}}}, \bibinfo {author} {\bibfnamefont {S.}~\bibnamefont {{Schulze}}}, \bibinfo {author} {\bibfnamefont {H.}~\bibnamefont {{Kumar}}}, \bibinfo {author} {\bibfnamefont {G.~C.}\ \bibnamefont {{Anupama}}}, \bibinfo {author} {\bibfnamefont {S.}~\bibnamefont {{Barway}}}, \bibinfo {author} {\bibfnamefont {E.~C.}\ \bibnamefont {{Bellm}}}, \bibinfo {author} {\bibfnamefont {V.}~\bibnamefont {{Bhalerao}}}, \bibinfo {author} {\bibfnamefont {T.-W.}\ \bibnamefont {{Chen}}}, \bibinfo {author} {\bibfnamefont {D.~A.}\ \bibnamefont {{Duev}}}, \bibinfo {author} {\bibfnamefont {L.}~\bibnamefont {{Galbany}}}, \bibinfo {author} {\bibfnamefont {M.~J.}\ \bibnamefont {{Graham}}}, \bibinfo
  {author} {\bibfnamefont {M.}~\bibnamefont {{Gromadzki}}}, \bibinfo {author} {\bibfnamefont {C.~P.}\ \bibnamefont {{Guti{\'e}rrez}}}, \bibinfo {author} {\bibfnamefont {N.}~\bibnamefont {{Ihanec}}}, \bibinfo {author} {\bibfnamefont {C.}~\bibnamefont {{Inserram}}}, \bibinfo {author} {\bibfnamefont {M.~M.}\ \bibnamefont {{Kasliwal}}}, \bibinfo {author} {\bibfnamefont {E.~C.}\ \bibnamefont {{Kool}}}, \bibinfo {author} {\bibfnamefont {S.~R.}\ \bibnamefont {{Kulkarni}}}, \bibinfo {author} {\bibfnamefont {R.~R.}\ \bibnamefont {{Laher}}}, \bibinfo {author} {\bibfnamefont {F.~J.}\ \bibnamefont {{Masci}}}, \bibinfo {author} {\bibfnamefont {J.~D.}\ \bibnamefont {{Neill}}}, \bibinfo {author} {\bibfnamefont {M.}~\bibnamefont {{Nicholl}}}, \bibinfo {author} {\bibfnamefont {M.}~\bibnamefont {{Pursiainen}}}, \bibinfo {author} {\bibfnamefont {J.}~\bibnamefont {{van Roestel}}}, \bibinfo {author} {\bibfnamefont {Y.}~\bibnamefont {{Sharma}}}, \bibinfo {author} {\bibfnamefont {J.}~\bibnamefont {{Sollerman}}}, \bibinfo {author}
  {\bibfnamefont {R.}~\bibnamefont {{Walters}}}, \ and\ \bibinfo {author} {\bibfnamefont {P.}~\bibnamefont {{Wiseman}}},\ }\href@noop {} {\bibfield  {journal} {\bibinfo  {journal} {arXiv e-prints}\ ,\ \bibinfo {eid} {arXiv:2103.01968}} (\bibinfo {year} {2021})},\ \Eprint {http://arxiv.org/abs/2103.01968} {arXiv:2103.01968 [astro-ph.HE]} \BibitemShut {NoStop}%
\bibitem [{\citenamefont {{Nayana}}\ and\ \citenamefont {{Chandra}}(2021)}]{Nayana+Chandra2021ApJ_18cow_Radio}%
  \BibitemOpen
  \bibfield  {author} {\bibinfo {author} {\bibfnamefont {A.~J.}\ \bibnamefont {{Nayana}}}\ and\ \bibinfo {author} {\bibfnamefont {P.}~\bibnamefont {{Chandra}}},\ }\href {\doibase 10.3847/2041-8213/abed55} {\bibfield  {journal} {\bibinfo  {journal} {\apjl}\ }\textbf {\bibinfo {volume} {912}},\ \bibinfo {eid} {L9} (\bibinfo {year} {2021})},\ \Eprint {http://arxiv.org/abs/2103.06008} {arXiv:2103.06008 [astro-ph.HE]} \BibitemShut {NoStop}%
\bibitem [{\citenamefont {{Ho}}\ \emph {et~al.}(2022)\citenamefont {{Ho}}, \citenamefont {{Margalit}}, \citenamefont {{Bremer}}, \citenamefont {{Perley}}, \citenamefont {{Yao}}, \citenamefont {{Dobie}}, \citenamefont {{Kaplan}}, \citenamefont {{O'Brien}}, \citenamefont {{Petitpas}},\ and\ \citenamefont {{Zic}}}]{Ho+2022ApJ_18cow_2020xnd_Radio}%
  \BibitemOpen
  \bibfield  {author} {\bibinfo {author} {\bibfnamefont {A.~Y.~Q.}\ \bibnamefont {{Ho}}}, \bibinfo {author} {\bibfnamefont {B.}~\bibnamefont {{Margalit}}}, \bibinfo {author} {\bibfnamefont {M.}~\bibnamefont {{Bremer}}}, \bibinfo {author} {\bibfnamefont {D.~A.}\ \bibnamefont {{Perley}}}, \bibinfo {author} {\bibfnamefont {Y.}~\bibnamefont {{Yao}}}, \bibinfo {author} {\bibfnamefont {D.}~\bibnamefont {{Dobie}}}, \bibinfo {author} {\bibfnamefont {D.~L.}\ \bibnamefont {{Kaplan}}}, \bibinfo {author} {\bibfnamefont {A.}~\bibnamefont {{O'Brien}}}, \bibinfo {author} {\bibfnamefont {G.}~\bibnamefont {{Petitpas}}}, \ and\ \bibinfo {author} {\bibfnamefont {A.}~\bibnamefont {{Zic}}},\ }\href {\doibase 10.3847/1538-4357/ac4e97} {\bibfield  {journal} {\bibinfo  {journal} {\apj}\ }\textbf {\bibinfo {volume} {932}},\ \bibinfo {eid} {116} (\bibinfo {year} {2022})},\ \Eprint {http://arxiv.org/abs/2110.05490} {arXiv:2110.05490 [astro-ph.HE]} \BibitemShut {NoStop}%
\bibitem [{\citenamefont {{Metzger}}\ and\ \citenamefont {{Perley}}(2023)}]{Metzger+Perley2023ApJ_FBOT_18cow_DustEchos}%
  \BibitemOpen
  \bibfield  {author} {\bibinfo {author} {\bibfnamefont {B.~D.}\ \bibnamefont {{Metzger}}}\ and\ \bibinfo {author} {\bibfnamefont {D.~A.}\ \bibnamefont {{Perley}}},\ }\href {\doibase 10.3847/1538-4357/acae89} {\bibfield  {journal} {\bibinfo  {journal} {\apj}\ }\textbf {\bibinfo {volume} {944}},\ \bibinfo {eid} {74} (\bibinfo {year} {2023})},\ \Eprint {http://arxiv.org/abs/2210.01819} {arXiv:2210.01819 [astro-ph.HE]} \BibitemShut {NoStop}%
\bibitem [{\citenamefont {{Ofek}}\ \emph {et~al.}(2023{\natexlab{a}})\citenamefont {{Ofek}}, \citenamefont {{Ben-Ami}}, \citenamefont {{Polishook}}, \citenamefont {{Segre}}, \citenamefont {{Blumenzweig}} \emph {et~al.}}]{Ofek+2023PASP_LAST_Overview}%
  \BibitemOpen
  \bibfield  {author} {\bibinfo {author} {\bibfnamefont {E.~O.}\ \bibnamefont {{Ofek}}}, \bibinfo {author} {\bibfnamefont {S.}~\bibnamefont {{Ben-Ami}}}, \bibinfo {author} {\bibfnamefont {D.}~\bibnamefont {{Polishook}}}, \bibinfo {author} {\bibfnamefont {E.}~\bibnamefont {{Segre}}}, \bibinfo {author} {\bibfnamefont {A.}~\bibnamefont {{Blumenzweig}}},  \emph {et~al.},\ }\href {\doibase 10.1088/1538-3873/acd8f0} {\bibfield  {journal} {\bibinfo  {journal} {\pasp}\ }\textbf {\bibinfo {volume} {135}},\ \bibinfo {eid} {065001} (\bibinfo {year} {2023}{\natexlab{a}})},\ \Eprint {http://arxiv.org/abs/2304.04796} {arXiv:2304.04796 [astro-ph.IM]} \BibitemShut {NoStop}%
\bibitem [{\citenamefont {{Ben-Ami}}\ \emph {et~al.}(2023)\citenamefont {{Ben-Ami}}, \citenamefont {{Ofek}}, \citenamefont {{Polishook}}, \citenamefont {{Franckowiak}}, \citenamefont {{Hallakoun}} \emph {et~al.}}]{BenAmi+2023PASP_LAST_Science}%
  \BibitemOpen
  \bibfield  {author} {\bibinfo {author} {\bibfnamefont {S.}~\bibnamefont {{Ben-Ami}}}, \bibinfo {author} {\bibfnamefont {E.~O.}\ \bibnamefont {{Ofek}}}, \bibinfo {author} {\bibfnamefont {D.}~\bibnamefont {{Polishook}}}, \bibinfo {author} {\bibfnamefont {A.}~\bibnamefont {{Franckowiak}}}, \bibinfo {author} {\bibfnamefont {N.}~\bibnamefont {{Hallakoun}}},  \emph {et~al.},\ }\href {\doibase 10.48550/arXiv.2304.02719} {\bibfield  {journal} {\bibinfo  {journal} {arXiv e-prints}\ ,\ \bibinfo {eid} {arXiv:2304.02719}} (\bibinfo {year} {2023})},\ \Eprint {http://arxiv.org/abs/2304.02719} {arXiv:2304.02719 [astro-ph.IM]} \BibitemShut {NoStop}%
\bibitem [{\citenamefont {{Bellm}}\ \emph {et~al.}(2019)\citenamefont {{Bellm}}, \citenamefont {{Kulkarni}}, \citenamefont {{Graham}}, \citenamefont {{Dekany}}, \citenamefont {{Smith}} \emph {et~al.}}]{Bellm+2019_ZTF_Overview}%
  \BibitemOpen
  \bibfield  {author} {\bibinfo {author} {\bibfnamefont {E.~C.}\ \bibnamefont {{Bellm}}}, \bibinfo {author} {\bibfnamefont {S.~R.}\ \bibnamefont {{Kulkarni}}}, \bibinfo {author} {\bibfnamefont {M.~J.}\ \bibnamefont {{Graham}}}, \bibinfo {author} {\bibfnamefont {R.}~\bibnamefont {{Dekany}}}, \bibinfo {author} {\bibfnamefont {R.~M.}\ \bibnamefont {{Smith}}},  \emph {et~al.},\ }\href {\doibase 10.1088/1538-3873/aaecbe} {\bibfield  {journal} {\bibinfo  {journal} {\pasp}\ }\textbf {\bibinfo {volume} {131}},\ \bibinfo {pages} {018002} (\bibinfo {year} {2019})},\ \Eprint {http://arxiv.org/abs/1902.01932} {arXiv:1902.01932 [astro-ph.IM]} \BibitemShut {NoStop}%
\bibitem [{\citenamefont {{Graham}}\ \emph {et~al.}(2019)\citenamefont {{Graham}}, \citenamefont {{Kulkarni}}, \citenamefont {{Bellm}}, \citenamefont {{Adams}}, \citenamefont {{Barbarino}}, \citenamefont {{Blagorodnova}}, \citenamefont {{Bodewits}}, \citenamefont {{Bolin}}, \citenamefont {{Brady}},\ and\ \citenamefont {{Cenko}}}]{Graham+2019_ZTF_objectives}%
  \BibitemOpen
  \bibfield  {author} {\bibinfo {author} {\bibfnamefont {M.~J.}\ \bibnamefont {{Graham}}}, \bibinfo {author} {\bibfnamefont {S.~R.}\ \bibnamefont {{Kulkarni}}}, \bibinfo {author} {\bibfnamefont {E.~C.}\ \bibnamefont {{Bellm}}}, \bibinfo {author} {\bibfnamefont {S.~M.}\ \bibnamefont {{Adams}}}, \bibinfo {author} {\bibfnamefont {C.}~\bibnamefont {{Barbarino}}}, \bibinfo {author} {\bibfnamefont {N.}~\bibnamefont {{Blagorodnova}}}, \bibinfo {author} {\bibfnamefont {D.}~\bibnamefont {{Bodewits}}}, \bibinfo {author} {\bibfnamefont {B.}~\bibnamefont {{Bolin}}}, \bibinfo {author} {\bibfnamefont {P.~R.}\ \bibnamefont {{Brady}}}, \ and\ \bibinfo {author} {\bibfnamefont {S.~B.}\ \bibnamefont {{Cenko}}},\ }\href {\doibase 10.1088/1538-3873/ab006c} {\bibfield  {journal} {\bibinfo  {journal} {\pasp}\ }\textbf {\bibinfo {volume} {131}},\ \bibinfo {pages} {078001} (\bibinfo {year} {2019})},\ \Eprint {http://arxiv.org/abs/1902.01945} {arXiv:1902.01945 [astro-ph.IM]} \BibitemShut {NoStop}%
\bibitem [{\citenamefont {{Masci}}\ \emph {et~al.}(2019)\citenamefont {{Masci}}, \citenamefont {{Laher}}, \citenamefont {{Rusholme}}, \citenamefont {{Shupe}}, \citenamefont {{Groom}}, \citenamefont {{Surace}}, \citenamefont {{Jackson}}, \citenamefont {{Monkewitz}}, \citenamefont {{Beck}},\ and\ \citenamefont {{Flynn}}}]{Masci+2019_ZTF_Pipeline}%
  \BibitemOpen
  \bibfield  {author} {\bibinfo {author} {\bibfnamefont {F.~J.}\ \bibnamefont {{Masci}}}, \bibinfo {author} {\bibfnamefont {R.~R.}\ \bibnamefont {{Laher}}}, \bibinfo {author} {\bibfnamefont {B.}~\bibnamefont {{Rusholme}}}, \bibinfo {author} {\bibfnamefont {D.~L.}\ \bibnamefont {{Shupe}}}, \bibinfo {author} {\bibfnamefont {S.}~\bibnamefont {{Groom}}}, \bibinfo {author} {\bibfnamefont {J.}~\bibnamefont {{Surace}}}, \bibinfo {author} {\bibfnamefont {E.}~\bibnamefont {{Jackson}}}, \bibinfo {author} {\bibfnamefont {S.}~\bibnamefont {{Monkewitz}}}, \bibinfo {author} {\bibfnamefont {R.}~\bibnamefont {{Beck}}}, \ and\ \bibinfo {author} {\bibfnamefont {D.}~\bibnamefont {{Flynn}}},\ }\href {\doibase 10.1088/1538-3873/aae8ac} {\bibfield  {journal} {\bibinfo  {journal} {\pasp}\ }\textbf {\bibinfo {volume} {131}},\ \bibinfo {pages} {018003} (\bibinfo {year} {2019})},\ \Eprint {http://arxiv.org/abs/1902.01872} {arXiv:1902.01872 [astro-ph.IM]} \BibitemShut {NoStop}%
\bibitem [{\citenamefont {{Ho}}\ \emph {et~al.}(2024)\citenamefont {{Ho}}, \citenamefont {{Srinivasaragavan}}, \citenamefont {{Perley}}, \citenamefont {{Andreoni}}, \citenamefont {{Rehentulla}},\ and\ \citenamefont {{Qin}}}]{Ho+2024TNSAN_AT2024wpp_discovery}%
  \BibitemOpen
  \bibfield  {author} {\bibinfo {author} {\bibfnamefont {A.~Y.~Q.}\ \bibnamefont {{Ho}}}, \bibinfo {author} {\bibfnamefont {G.}~\bibnamefont {{Srinivasaragavan}}}, \bibinfo {author} {\bibfnamefont {D.}~\bibnamefont {{Perley}}}, \bibinfo {author} {\bibfnamefont {I.}~\bibnamefont {{Andreoni}}}, \bibinfo {author} {\bibfnamefont {N.}~\bibnamefont {{Rehentulla}}}, \ and\ \bibinfo {author} {\bibfnamefont {Y.}~\bibnamefont {{Qin}}},\ }\href@noop {} {\bibfield  {journal} {\bibinfo  {journal} {Transient Name Server AstroNote}\ }\textbf {\bibinfo {volume} {272}},\ \bibinfo {pages} {1} (\bibinfo {year} {2024})}\BibitemShut {NoStop}%
\bibitem [{\citenamefont {{Sfaradi}}\ \emph {et~al.}(2024)\citenamefont {{Sfaradi}}, \citenamefont {{Margutti}}, \citenamefont {{Farah}}, \citenamefont {{Wiston}}, \citenamefont {{J}}, \citenamefont {{Bright}}, \citenamefont {{Chornock}}, \citenamefont {{LeBaron}}, \citenamefont {{Hammerstein}}, \citenamefont {{Brethauer}}, \citenamefont {{Laskar}}, \citenamefont {{Siemion}}, \citenamefont {{Pollak}}, \citenamefont {{Sheikh}}, \citenamefont {{Sears}},\ and\ \citenamefont {{Migliori}}}]{Sfaradi+2024TNSAN_AT2024wpp_redshift}%
  \BibitemOpen
  \bibfield  {author} {\bibinfo {author} {\bibfnamefont {I.}~\bibnamefont {{Sfaradi}}}, \bibinfo {author} {\bibfnamefont {R.}~\bibnamefont {{Margutti}}}, \bibinfo {author} {\bibfnamefont {W.}~\bibnamefont {{Farah}}}, \bibinfo {author} {\bibfnamefont {E.}~\bibnamefont {{Wiston}}}, \bibinfo {author} {\bibfnamefont {N.~A.}\ \bibnamefont {{J}}}, \bibinfo {author} {\bibfnamefont {J.}~\bibnamefont {{Bright}}}, \bibinfo {author} {\bibfnamefont {R.}~\bibnamefont {{Chornock}}}, \bibinfo {author} {\bibfnamefont {N.}~\bibnamefont {{LeBaron}}}, \bibinfo {author} {\bibfnamefont {E.}~\bibnamefont {{Hammerstein}}}, \bibinfo {author} {\bibfnamefont {D.}~\bibnamefont {{Brethauer}}}, \bibinfo {author} {\bibfnamefont {T.}~\bibnamefont {{Laskar}}}, \bibinfo {author} {\bibfnamefont {A.}~\bibnamefont {{Siemion}}}, \bibinfo {author} {\bibfnamefont {A.}~\bibnamefont {{Pollak}}}, \bibinfo {author} {\bibfnamefont {S.}~\bibnamefont {{Sheikh}}}, \bibinfo {author} {\bibfnamefont {H.}~\bibnamefont {{Sears}}}, \ and\ \bibinfo {author}
  {\bibfnamefont {G.}~\bibnamefont {{Migliori}}},\ }\href@noop {} {\bibfield  {journal} {\bibinfo  {journal} {Transient Name Server AstroNote}\ }\textbf {\bibinfo {volume} {290}},\ \bibinfo {pages} {1} (\bibinfo {year} {2024})}\BibitemShut {NoStop}%
\bibitem [{\citenamefont {{Perley}}\ \emph {et~al.}(2024)\citenamefont {{Perley}}, \citenamefont {{Qin}}, \citenamefont {{Rich}}, \citenamefont {{Daddi}}, \citenamefont {{Collins}}, \citenamefont {{Gatkine}}, \citenamefont {{Neill}}, \citenamefont {{Hinds}},\ and\ \citenamefont {{McGurk}}}]{Perley+2024TNSAN_AT2024wpp_Redshift}%
  \BibitemOpen
  \bibfield  {author} {\bibinfo {author} {\bibfnamefont {D.~A.}\ \bibnamefont {{Perley}}}, \bibinfo {author} {\bibfnamefont {Y.}~\bibnamefont {{Qin}}}, \bibinfo {author} {\bibfnamefont {R.~M.}\ \bibnamefont {{Rich}}}, \bibinfo {author} {\bibfnamefont {E.}~\bibnamefont {{Daddi}}}, \bibinfo {author} {\bibfnamefont {M.}~\bibnamefont {{Collins}}}, \bibinfo {author} {\bibfnamefont {P.}~\bibnamefont {{Gatkine}}}, \bibinfo {author} {\bibfnamefont {D.}~\bibnamefont {{Neill}}}, \bibinfo {author} {\bibfnamefont {K.}~\bibnamefont {{Hinds}}}, \ and\ \bibinfo {author} {\bibfnamefont {R.}~\bibnamefont {{McGurk}}},\ }\href@noop {} {\bibfield  {journal} {\bibinfo  {journal} {Transient Name Server AstroNote}\ }\textbf {\bibinfo {volume} {280}},\ \bibinfo {pages} {1} (\bibinfo {year} {2024})}\BibitemShut {NoStop}%
\bibitem [{\citenamefont {{Schroeder}}\ \emph {et~al.}(2024)\citenamefont {{Schroeder}}, \citenamefont {{Ho}},\ and\ \citenamefont {{Perley}}}]{Schroeder+2024TNSAN_AT2024wpp_VLA}%
  \BibitemOpen
  \bibfield  {author} {\bibinfo {author} {\bibfnamefont {G.}~\bibnamefont {{Schroeder}}}, \bibinfo {author} {\bibfnamefont {A.~Y.~Q.}\ \bibnamefont {{Ho}}}, \ and\ \bibinfo {author} {\bibfnamefont {D.~A.}\ \bibnamefont {{Perley}}},\ }\href@noop {} {\bibfield  {journal} {\bibinfo  {journal} {Transient Name Server AstroNote}\ }\textbf {\bibinfo {volume} {314}},\ \bibinfo {pages} {1} (\bibinfo {year} {2024})}\BibitemShut {NoStop}%
\bibitem [{\citenamefont {{Gehrels}}\ \emph {et~al.}(2004)\citenamefont {{Gehrels}}, \citenamefont {{Chincarini}}, \citenamefont {{Giommi}}, \citenamefont {{Mason}}, \citenamefont {{Nousek}}, \citenamefont {{Wells}}, \citenamefont {{White}}, \citenamefont {{Barthelmy}}, \citenamefont {{Burrows}},\ and\ \citenamefont {{Cominsky}}}]{Gehrels+2004_Swift}%
  \BibitemOpen
  \bibfield  {author} {\bibinfo {author} {\bibfnamefont {N.}~\bibnamefont {{Gehrels}}}, \bibinfo {author} {\bibfnamefont {G.}~\bibnamefont {{Chincarini}}}, \bibinfo {author} {\bibfnamefont {P.}~\bibnamefont {{Giommi}}}, \bibinfo {author} {\bibfnamefont {K.~O.}\ \bibnamefont {{Mason}}}, \bibinfo {author} {\bibfnamefont {J.~A.}\ \bibnamefont {{Nousek}}}, \bibinfo {author} {\bibfnamefont {A.~A.}\ \bibnamefont {{Wells}}}, \bibinfo {author} {\bibfnamefont {N.~E.}\ \bibnamefont {{White}}}, \bibinfo {author} {\bibfnamefont {S.~D.}\ \bibnamefont {{Barthelmy}}}, \bibinfo {author} {\bibfnamefont {D.~N.}\ \bibnamefont {{Burrows}}}, \ and\ \bibinfo {author} {\bibfnamefont {L.~R.}\ \bibnamefont {{Cominsky}}},\ }\href {\doibase 10.1086/422091} {\bibfield  {journal} {\bibinfo  {journal} {\apj}\ }\textbf {\bibinfo {volume} {611}},\ \bibinfo {pages} {1005} (\bibinfo {year} {2004})},\ \Eprint {http://arxiv.org/abs/astro-ph/0405233} {arXiv:astro-ph/0405233 [astro-ph]} \BibitemShut {NoStop}%
\bibitem [{\citenamefont {{Srinivasaragavan}}\ \emph {et~al.}(2024)\citenamefont {{Srinivasaragavan}}, \citenamefont {{Ho}}, \citenamefont {{Perley}}, \citenamefont {{Andreoni}}, \citenamefont {{Rehentulla}}, \citenamefont {{Qin}},\ and\ \citenamefont {{Bellm}}}]{Srinivasaragavan+2024TNSAN_AT2024wpp_Xray}%
  \BibitemOpen
  \bibfield  {author} {\bibinfo {author} {\bibfnamefont {G.}~\bibnamefont {{Srinivasaragavan}}}, \bibinfo {author} {\bibfnamefont {A.}~\bibnamefont {{Ho}}}, \bibinfo {author} {\bibfnamefont {D.}~\bibnamefont {{Perley}}}, \bibinfo {author} {\bibfnamefont {I.}~\bibnamefont {{Andreoni}}}, \bibinfo {author} {\bibfnamefont {N.}~\bibnamefont {{Rehentulla}}}, \bibinfo {author} {\bibfnamefont {Y.}~\bibnamefont {{Qin}}}, \ and\ \bibinfo {author} {\bibfnamefont {E.}~\bibnamefont {{Bellm}}},\ }\href@noop {} {\bibfield  {journal} {\bibinfo  {journal} {Transient Name Server AstroNote}\ }\textbf {\bibinfo {volume} {276}},\ \bibinfo {pages} {1} (\bibinfo {year} {2024})}\BibitemShut {NoStop}%
\bibitem [{\citenamefont {{Margutti}}\ \emph {et~al.}(2024)\citenamefont {{Margutti}}, \citenamefont {{J}}, \citenamefont {{Chornock}}, \citenamefont {{Guo}}, \citenamefont {{LeBaron}}, \citenamefont {{Wiston}}, \citenamefont {{Sfaradi}}, \citenamefont {{Hammerstein}}, \citenamefont {{Brethauer}}, \citenamefont {{Sears}}, \citenamefont {{Migliori}}, \citenamefont {{Laskar}}, \citenamefont {{Metzger}}, \citenamefont {{Pasham}}, \citenamefont {{Aspegren}}, \citenamefont {{Lu}},\ and\ \citenamefont {{Milisavljevic}}}]{Margutti+2024TNSAN_AT2024wpp_Xray}%
  \BibitemOpen
  \bibfield  {author} {\bibinfo {author} {\bibfnamefont {R.}~\bibnamefont {{Margutti}}}, \bibinfo {author} {\bibfnamefont {N.~A.}\ \bibnamefont {{J}}}, \bibinfo {author} {\bibfnamefont {R.}~\bibnamefont {{Chornock}}}, \bibinfo {author} {\bibfnamefont {X.}~\bibnamefont {{Guo}}}, \bibinfo {author} {\bibfnamefont {N.}~\bibnamefont {{LeBaron}}}, \bibinfo {author} {\bibfnamefont {E.}~\bibnamefont {{Wiston}}}, \bibinfo {author} {\bibfnamefont {I.}~\bibnamefont {{Sfaradi}}}, \bibinfo {author} {\bibfnamefont {E.}~\bibnamefont {{Hammerstein}}}, \bibinfo {author} {\bibfnamefont {D.}~\bibnamefont {{Brethauer}}}, \bibinfo {author} {\bibfnamefont {H.}~\bibnamefont {{Sears}}}, \bibinfo {author} {\bibfnamefont {G.}~\bibnamefont {{Migliori}}}, \bibinfo {author} {\bibfnamefont {T.}~\bibnamefont {{Laskar}}}, \bibinfo {author} {\bibfnamefont {B.~D.}\ \bibnamefont {{Metzger}}}, \bibinfo {author} {\bibfnamefont {D.}~\bibnamefont {{Pasham}}}, \bibinfo {author} {\bibfnamefont {O.}~\bibnamefont {{Aspegren}}}, \bibinfo {author}
  {\bibfnamefont {W.}~\bibnamefont {{Lu}}}, \ and\ \bibinfo {author} {\bibfnamefont {D.}~\bibnamefont {{Milisavljevic}}},\ }\href@noop {} {\bibfield  {journal} {\bibinfo  {journal} {Transient Name Server AstroNote}\ }\textbf {\bibinfo {volume} {339}},\ \bibinfo {pages} {1} (\bibinfo {year} {2024})}\BibitemShut {NoStop}%
\bibitem [{\citenamefont {{Pursiainen}}\ \emph {et~al.}(2025)\citenamefont {{Pursiainen}}, \citenamefont {{Killestein}}, \citenamefont {{Kuncarayakti}}, \citenamefont {{Charalampopoulos}}, \citenamefont {{Warwick}}, \citenamefont {{Lyman}}, \citenamefont {{Kotak}}, \citenamefont {{Leloudas}}, \citenamefont {{Coppejans}}, \citenamefont {{Kravtsov}}, \citenamefont {{Maeda}}, \citenamefont {{Nagao}}, \citenamefont {{Taguchi}}, \citenamefont {{Ackley}}, \citenamefont {{Dhillon}}, \citenamefont {{Galloway}}, \citenamefont {{Kumar}}, \citenamefont {{O'Neill}}, \citenamefont {{Ramsay}},\ and\ \citenamefont {{Steeghs}}}]{Pursiainen+2025MNRAS_AT2024wpp_spherical}%
  \BibitemOpen
  \bibfield  {author} {\bibinfo {author} {\bibfnamefont {M.}~\bibnamefont {{Pursiainen}}}, \bibinfo {author} {\bibfnamefont {T.~L.}\ \bibnamefont {{Killestein}}}, \bibinfo {author} {\bibfnamefont {H.}~\bibnamefont {{Kuncarayakti}}}, \bibinfo {author} {\bibfnamefont {P.}~\bibnamefont {{Charalampopoulos}}}, \bibinfo {author} {\bibfnamefont {B.}~\bibnamefont {{Warwick}}}, \bibinfo {author} {\bibfnamefont {J.}~\bibnamefont {{Lyman}}}, \bibinfo {author} {\bibfnamefont {R.}~\bibnamefont {{Kotak}}}, \bibinfo {author} {\bibfnamefont {G.}~\bibnamefont {{Leloudas}}}, \bibinfo {author} {\bibfnamefont {D.}~\bibnamefont {{Coppejans}}}, \bibinfo {author} {\bibfnamefont {T.}~\bibnamefont {{Kravtsov}}}, \bibinfo {author} {\bibfnamefont {K.}~\bibnamefont {{Maeda}}}, \bibinfo {author} {\bibfnamefont {T.}~\bibnamefont {{Nagao}}}, \bibinfo {author} {\bibfnamefont {K.}~\bibnamefont {{Taguchi}}}, \bibinfo {author} {\bibfnamefont {K.}~\bibnamefont {{Ackley}}}, \bibinfo {author} {\bibfnamefont {V.~S.}\ \bibnamefont {{Dhillon}}},
  \bibinfo {author} {\bibfnamefont {D.~K.}\ \bibnamefont {{Galloway}}}, \bibinfo {author} {\bibfnamefont {A.}~\bibnamefont {{Kumar}}}, \bibinfo {author} {\bibfnamefont {D.}~\bibnamefont {{O'Neill}}}, \bibinfo {author} {\bibfnamefont {G.}~\bibnamefont {{Ramsay}}}, \ and\ \bibinfo {author} {\bibfnamefont {D.}~\bibnamefont {{Steeghs}}},\ }\href {\doibase 10.1093/mnras/staf232} {\bibfield  {journal} {\bibinfo  {journal} {\mnras}\ }\textbf {\bibinfo {volume} {537}},\ \bibinfo {pages} {3298} (\bibinfo {year} {2025})},\ \Eprint {http://arxiv.org/abs/2411.03272} {arXiv:2411.03272 [astro-ph.HE]} \BibitemShut {NoStop}%
\bibitem [{\citenamefont {{Maund}}\ \emph {et~al.}(2023)\citenamefont {{Maund}}, \citenamefont {{H{\"o}flich}}, \citenamefont {{Steele}}, \citenamefont {{Yang}}, \citenamefont {{Wiersema}}, \citenamefont {{Kobayashi}}, \citenamefont {{Jordana-Mitjans}}, \citenamefont {{Mundell}}, \citenamefont {{Gomboc}}, \citenamefont {{Guidorzi}},\ and\ \citenamefont {{Smith}}}]{Maund+2023MNRAS_AT2018cow_polarization}%
  \BibitemOpen
  \bibfield  {author} {\bibinfo {author} {\bibfnamefont {J.~R.}\ \bibnamefont {{Maund}}}, \bibinfo {author} {\bibfnamefont {P.~A.}\ \bibnamefont {{H{\"o}flich}}}, \bibinfo {author} {\bibfnamefont {I.~A.}\ \bibnamefont {{Steele}}}, \bibinfo {author} {\bibfnamefont {Y.}~\bibnamefont {{Yang}}}, \bibinfo {author} {\bibfnamefont {K.}~\bibnamefont {{Wiersema}}}, \bibinfo {author} {\bibfnamefont {S.}~\bibnamefont {{Kobayashi}}}, \bibinfo {author} {\bibfnamefont {N.}~\bibnamefont {{Jordana-Mitjans}}}, \bibinfo {author} {\bibfnamefont {C.}~\bibnamefont {{Mundell}}}, \bibinfo {author} {\bibfnamefont {A.}~\bibnamefont {{Gomboc}}}, \bibinfo {author} {\bibfnamefont {C.}~\bibnamefont {{Guidorzi}}}, \ and\ \bibinfo {author} {\bibfnamefont {R.~J.}\ \bibnamefont {{Smith}}},\ }\href {\doibase 10.1093/mnras/stad539} {\bibfield  {journal} {\bibinfo  {journal} {\mnras}\ }\textbf {\bibinfo {volume} {521}},\ \bibinfo {pages} {3323} (\bibinfo {year} {2023})},\ \Eprint {http://arxiv.org/abs/2303.00787} {arXiv:2303.00787
  [astro-ph.SR]} \BibitemShut {NoStop}%
\bibitem [{\citenamefont {{Ofek}}\ \emph {et~al.}(2023{\natexlab{b}})\citenamefont {{Ofek}}, \citenamefont {{Shvartzvald}}, \citenamefont {{Sharon}}, \citenamefont {{Tishler}}, \citenamefont {{Elhanati}} \emph {et~al.}}]{Ofek+2023PASP_LAST_PipeplineI}%
  \BibitemOpen
  \bibfield  {author} {\bibinfo {author} {\bibfnamefont {E.~O.}\ \bibnamefont {{Ofek}}}, \bibinfo {author} {\bibfnamefont {Y.}~\bibnamefont {{Shvartzvald}}}, \bibinfo {author} {\bibfnamefont {A.}~\bibnamefont {{Sharon}}}, \bibinfo {author} {\bibfnamefont {C.}~\bibnamefont {{Tishler}}}, \bibinfo {author} {\bibfnamefont {D.}~\bibnamefont {{Elhanati}}},  \emph {et~al.},\ }\href {\doibase 10.1088/1538-3873/ad0977} {\bibfield  {journal} {\bibinfo  {journal} {\pasp}\ }\textbf {\bibinfo {volume} {135}},\ \bibinfo {eid} {124502} (\bibinfo {year} {2023}{\natexlab{b}})},\ \Eprint {http://arxiv.org/abs/2310.13063} {arXiv:2310.13063 [astro-ph.IM]} \BibitemShut {NoStop}%
\bibitem [{\citenamefont {{Ofek}}\ and\ \citenamefont {{Ben-Ami}}(2020)}]{Ofek+BenAmi2020_Grasp_SkySurvrys_CostEffectivness}%
  \BibitemOpen
  \bibfield  {author} {\bibinfo {author} {\bibfnamefont {E.~O.}\ \bibnamefont {{Ofek}}}\ and\ \bibinfo {author} {\bibfnamefont {S.}~\bibnamefont {{Ben-Ami}}},\ }\href@noop {} {\bibfield  {journal} {\bibinfo  {journal} {arXiv e-prints}\ ,\ \bibinfo {eid} {arXiv:2011.04674}} (\bibinfo {year} {2020})},\ \Eprint {http://arxiv.org/abs/2011.04674} {arXiv:2011.04674 [astro-ph.IM]} \BibitemShut {NoStop}%
\bibitem [{\citenamefont {{Zackay}}\ and\ \citenamefont {{Ofek}}(2017{\natexlab{a}})}]{Zackay+2017_CoadditionI}%
  \BibitemOpen
  \bibfield  {author} {\bibinfo {author} {\bibfnamefont {B.}~\bibnamefont {{Zackay}}}\ and\ \bibinfo {author} {\bibfnamefont {E.~O.}\ \bibnamefont {{Ofek}}},\ }\href {\doibase 10.3847/1538-4357/836/2/187} {\bibfield  {journal} {\bibinfo  {journal} {\apj}\ }\textbf {\bibinfo {volume} {836}},\ \bibinfo {eid} {187} (\bibinfo {year} {2017}{\natexlab{a}})},\ \Eprint {http://arxiv.org/abs/1512.06872} {arXiv:1512.06872 [astro-ph.IM]} \BibitemShut {NoStop}%
\bibitem [{\citenamefont {{Gaia Collaboration}}\ \emph {et~al.}(2016)\citenamefont {{Gaia Collaboration}}, \citenamefont {{Prusti}}, \citenamefont {{de Bruijne}}, \citenamefont {{Brown}}, \citenamefont {{Vallenari}}, \citenamefont {{Babusiaux}}, \citenamefont {{Bailer-Jones}}, \citenamefont {{Bastian}}, \citenamefont {{Biermann}}, \citenamefont {{Evans}}, \citenamefont {{Eyer}}, \citenamefont {{Jansen}}, \citenamefont {{Jordi}}, \citenamefont {{Klioner}}, \citenamefont {{Lammers}}, \citenamefont {{Lindegren}}, \citenamefont {{Luri}}, \citenamefont {{Mignard}}, \citenamefont {{Milligan}},\ and\ \citenamefont {{Panem}}}]{GAIA+2016_GAIA_mission}%
  \BibitemOpen
  \bibfield  {author} {\bibinfo {author} {\bibnamefont {{Gaia Collaboration}}}, \bibinfo {author} {\bibfnamefont {T.}~\bibnamefont {{Prusti}}}, \bibinfo {author} {\bibfnamefont {J.~H.~J.}\ \bibnamefont {{de Bruijne}}}, \bibinfo {author} {\bibfnamefont {A.~G.~A.}\ \bibnamefont {{Brown}}}, \bibinfo {author} {\bibfnamefont {A.}~\bibnamefont {{Vallenari}}}, \bibinfo {author} {\bibfnamefont {C.}~\bibnamefont {{Babusiaux}}}, \bibinfo {author} {\bibfnamefont {C.~A.~L.}\ \bibnamefont {{Bailer-Jones}}}, \bibinfo {author} {\bibfnamefont {U.}~\bibnamefont {{Bastian}}}, \bibinfo {author} {\bibfnamefont {M.}~\bibnamefont {{Biermann}}}, \bibinfo {author} {\bibfnamefont {D.~W.}\ \bibnamefont {{Evans}}}, \bibinfo {author} {\bibfnamefont {L.}~\bibnamefont {{Eyer}}}, \bibinfo {author} {\bibfnamefont {F.}~\bibnamefont {{Jansen}}}, \bibinfo {author} {\bibfnamefont {C.}~\bibnamefont {{Jordi}}}, \bibinfo {author} {\bibfnamefont {S.~A.}\ \bibnamefont {{Klioner}}}, \bibinfo {author} {\bibfnamefont {U.}~\bibnamefont {{Lammers}}},
  \bibinfo {author} {\bibfnamefont {L.}~\bibnamefont {{Lindegren}}}, \bibinfo {author} {\bibfnamefont {X.}~\bibnamefont {{Luri}}}, \bibinfo {author} {\bibfnamefont {F.}~\bibnamefont {{Mignard}}}, \bibinfo {author} {\bibfnamefont {D.~J.}\ \bibnamefont {{Milligan}}}, \ and\ \bibinfo {author} {\bibfnamefont {C.~e.~a.}\ \bibnamefont {{Panem}}},\ }\href {\doibase 10.1051/0004-6361/201629272} {\bibfield  {journal} {\bibinfo  {journal} {\aap}\ }\textbf {\bibinfo {volume} {595}},\ \bibinfo {eid} {A1} (\bibinfo {year} {2016})},\ \Eprint {http://arxiv.org/abs/1609.04153} {arXiv:1609.04153 [astro-ph.IM]} \BibitemShut {NoStop}%
\bibitem [{\citenamefont {{Gaia Collaboration}}(2022)}]{GAIA+2022yCat_GAIA_DR3_MainSourcesCatalog}%
  \BibitemOpen
  \bibfield  {author} {\bibinfo {author} {\bibnamefont {{Gaia Collaboration}}},\ }\href@noop {} {\bibfield  {journal} {\bibinfo  {journal} {VizieR Online Data Catalog}\ ,\ \bibinfo {eid} {I/355}} (\bibinfo {year} {2022})}\BibitemShut {NoStop}%
\bibitem [{\citenamefont {{Zackay}}\ \emph {et~al.}(2016)\citenamefont {{Zackay}}, \citenamefont {{Ofek}},\ and\ \citenamefont {{Gal-Yam}}}]{Zackay+2016_ZOGY_ImageSubtraction}%
  \BibitemOpen
  \bibfield  {author} {\bibinfo {author} {\bibfnamefont {B.}~\bibnamefont {{Zackay}}}, \bibinfo {author} {\bibfnamefont {E.~O.}\ \bibnamefont {{Ofek}}}, \ and\ \bibinfo {author} {\bibfnamefont {A.}~\bibnamefont {{Gal-Yam}}},\ }\href {\doibase 10.3847/0004-637X/830/1/27} {\bibfield  {journal} {\bibinfo  {journal} {\apj}\ }\textbf {\bibinfo {volume} {830}},\ \bibinfo {eid} {27} (\bibinfo {year} {2016})},\ \Eprint {http://arxiv.org/abs/1601.02655} {arXiv:1601.02655 [astro-ph.IM]} \BibitemShut {NoStop}%
\bibitem [{\citenamefont {{Springer}}\ \emph {et~al.}(2024)\citenamefont {{Springer}}, \citenamefont {{Ofek}}, \citenamefont {{Zackay}}, \citenamefont {{Konno}}, \citenamefont {{Sharon}}, \citenamefont {{Nir}}, \citenamefont {{Rubin}}, \citenamefont {{Haddad}}, \citenamefont {{Friedman}}, \citenamefont {{Schein Lubomirsky}}, \citenamefont {{Aizenberg}}, \citenamefont {{Krassilchtchikov}},\ and\ \citenamefont {{Gal-Yam}}}]{Springer+Ofek+2024AJ_Translient}%
  \BibitemOpen
  \bibfield  {author} {\bibinfo {author} {\bibfnamefont {O.}~\bibnamefont {{Springer}}}, \bibinfo {author} {\bibfnamefont {E.~O.}\ \bibnamefont {{Ofek}}}, \bibinfo {author} {\bibfnamefont {B.}~\bibnamefont {{Zackay}}}, \bibinfo {author} {\bibfnamefont {R.}~\bibnamefont {{Konno}}}, \bibinfo {author} {\bibfnamefont {A.}~\bibnamefont {{Sharon}}}, \bibinfo {author} {\bibfnamefont {G.}~\bibnamefont {{Nir}}}, \bibinfo {author} {\bibfnamefont {A.}~\bibnamefont {{Rubin}}}, \bibinfo {author} {\bibfnamefont {A.}~\bibnamefont {{Haddad}}}, \bibinfo {author} {\bibfnamefont {J.}~\bibnamefont {{Friedman}}}, \bibinfo {author} {\bibfnamefont {L.}~\bibnamefont {{Schein Lubomirsky}}}, \bibinfo {author} {\bibfnamefont {I.}~\bibnamefont {{Aizenberg}}}, \bibinfo {author} {\bibfnamefont {A.}~\bibnamefont {{Krassilchtchikov}}}, \ and\ \bibinfo {author} {\bibfnamefont {A.}~\bibnamefont {{Gal-Yam}}},\ }\href {\doibase 10.48550/arXiv.2403.09771} {\bibfield  {journal} {\bibinfo  {journal} {arXiv e-prints}\ ,\ \bibinfo {eid}
  {arXiv:2403.09771}} (\bibinfo {year} {2024})},\ \Eprint {http://arxiv.org/abs/2403.09771} {arXiv:2403.09771 [astro-ph.IM]} \BibitemShut {NoStop}%
\bibitem [{\citenamefont {{Corbett}}\ \emph {et~al.}(2020)\citenamefont {{Corbett}}, \citenamefont {{Law}}, \citenamefont {{Soto}}, \citenamefont {{Howard}}, \citenamefont {{Glazier}}, \citenamefont {{Gonzalez}}, \citenamefont {{Ratzloff}}, \citenamefont {{Galliher}}, \citenamefont {{Fors}},\ and\ \citenamefont {{Quimby}}}]{Corbett+2020_SatellitesGlints}%
  \BibitemOpen
  \bibfield  {author} {\bibinfo {author} {\bibfnamefont {H.}~\bibnamefont {{Corbett}}}, \bibinfo {author} {\bibfnamefont {N.~M.}\ \bibnamefont {{Law}}}, \bibinfo {author} {\bibfnamefont {A.~V.}\ \bibnamefont {{Soto}}}, \bibinfo {author} {\bibfnamefont {W.~S.}\ \bibnamefont {{Howard}}}, \bibinfo {author} {\bibfnamefont {A.}~\bibnamefont {{Glazier}}}, \bibinfo {author} {\bibfnamefont {R.}~\bibnamefont {{Gonzalez}}}, \bibinfo {author} {\bibfnamefont {J.~K.}\ \bibnamefont {{Ratzloff}}}, \bibinfo {author} {\bibfnamefont {N.}~\bibnamefont {{Galliher}}}, \bibinfo {author} {\bibfnamefont {O.}~\bibnamefont {{Fors}}}, \ and\ \bibinfo {author} {\bibfnamefont {R.}~\bibnamefont {{Quimby}}},\ }\href {\doibase 10.3847/2041-8213/abbee5} {\bibfield  {journal} {\bibinfo  {journal} {\apjl}\ }\textbf {\bibinfo {volume} {903}},\ \bibinfo {eid} {L27} (\bibinfo {year} {2020})},\ \Eprint {http://arxiv.org/abs/2011.02495} {arXiv:2011.02495 [astro-ph.HE]} \BibitemShut {NoStop}%
\bibitem [{\citenamefont {{Nir}}\ \emph {et~al.}(2021{\natexlab{a}})\citenamefont {{Nir}}, \citenamefont {{Ofek}}, \citenamefont {{Ben-Ami}}, \citenamefont {{Segev}}, \citenamefont {{Polishook}},\ and\ \citenamefont {{Manulis}}}]{Nir+2020_Satellites_Glints_FlaresLimit}%
  \BibitemOpen
  \bibfield  {author} {\bibinfo {author} {\bibfnamefont {G.}~\bibnamefont {{Nir}}}, \bibinfo {author} {\bibfnamefont {E.~O.}\ \bibnamefont {{Ofek}}}, \bibinfo {author} {\bibfnamefont {S.}~\bibnamefont {{Ben-Ami}}}, \bibinfo {author} {\bibfnamefont {N.}~\bibnamefont {{Segev}}}, \bibinfo {author} {\bibfnamefont {D.}~\bibnamefont {{Polishook}}}, \ and\ \bibinfo {author} {\bibfnamefont {I.}~\bibnamefont {{Manulis}}},\ }\href {\doibase 10.1093/mnras/stab1437} {\bibfield  {journal} {\bibinfo  {journal} {\mnras}\ }\textbf {\bibinfo {volume} {505}},\ \bibinfo {pages} {2477} (\bibinfo {year} {2021}{\natexlab{a}})},\ \Eprint {http://arxiv.org/abs/2011.03497} {arXiv:2011.03497 [astro-ph.IM]} \BibitemShut {NoStop}%
\bibitem [{\citenamefont {{Nir}}\ \emph {et~al.}(2021{\natexlab{b}})\citenamefont {{Nir}}, \citenamefont {{Ofek}},\ and\ \citenamefont {{Gal-Yam}}}]{Nir+2021_RNASS_GN-z11-Flash_SatelliteGlint}%
  \BibitemOpen
  \bibfield  {author} {\bibinfo {author} {\bibfnamefont {G.}~\bibnamefont {{Nir}}}, \bibinfo {author} {\bibfnamefont {E.~O.}\ \bibnamefont {{Ofek}}}, \ and\ \bibinfo {author} {\bibfnamefont {A.}~\bibnamefont {{Gal-Yam}}},\ }\href {\doibase 10.3847/2515-5172/abe540} {\bibfield  {journal} {\bibinfo  {journal} {Research Notes of the American Astronomical Society}\ }\textbf {\bibinfo {volume} {5}},\ \bibinfo {eid} {27} (\bibinfo {year} {2021}{\natexlab{b}})},\ \Eprint {http://arxiv.org/abs/2102.04466} {arXiv:2102.04466 [astro-ph.HE]} \BibitemShut {NoStop}%
\bibitem [{\citenamefont {{Zackay}}\ and\ \citenamefont {{Ofek}}(2017{\natexlab{b}})}]{Zackay+2017_CoadditionII}%
  \BibitemOpen
  \bibfield  {author} {\bibinfo {author} {\bibfnamefont {B.}~\bibnamefont {{Zackay}}}\ and\ \bibinfo {author} {\bibfnamefont {E.~O.}\ \bibnamefont {{Ofek}}},\ }\href {\doibase 10.3847/1538-4357/836/2/188} {\bibfield  {journal} {\bibinfo  {journal} {\apj}\ }\textbf {\bibinfo {volume} {836}},\ \bibinfo {eid} {188} (\bibinfo {year} {2017}{\natexlab{b}})},\ \Eprint {http://arxiv.org/abs/1512.06879} {arXiv:1512.06879 [astro-ph.IM]} \BibitemShut {NoStop}%
\bibitem [{\citenamefont {{Hinshaw}}\ \emph {et~al.}(2013)\citenamefont {{Hinshaw}}, \citenamefont {{Larson}}, \citenamefont {{Komatsu}}, \citenamefont {{Spergel}}, \citenamefont {{Bennett}}, \citenamefont {{Dunkley}}, \citenamefont {{Nolta}}, \citenamefont {{Halpern}}, \citenamefont {{Hill}}, \citenamefont {{Odegard}}, \citenamefont {{Page}}, \citenamefont {{Smith}}, \citenamefont {{Weiland}}, \citenamefont {{Gold}}, \citenamefont {{Jarosik}}, \citenamefont {{Kogut}}, \citenamefont {{Limon}}, \citenamefont {{Meyer}}, \citenamefont {{Tucker}}, \citenamefont {{Wollack}},\ and\ \citenamefont {{Wright}}}]{Hinshaw+2013_WMAP_9yr_CosmologicalParameters}%
  \BibitemOpen
  \bibfield  {author} {\bibinfo {author} {\bibfnamefont {G.}~\bibnamefont {{Hinshaw}}}, \bibinfo {author} {\bibfnamefont {D.}~\bibnamefont {{Larson}}}, \bibinfo {author} {\bibfnamefont {E.}~\bibnamefont {{Komatsu}}}, \bibinfo {author} {\bibfnamefont {D.~N.}\ \bibnamefont {{Spergel}}}, \bibinfo {author} {\bibfnamefont {C.~L.}\ \bibnamefont {{Bennett}}}, \bibinfo {author} {\bibfnamefont {J.}~\bibnamefont {{Dunkley}}}, \bibinfo {author} {\bibfnamefont {M.~R.}\ \bibnamefont {{Nolta}}}, \bibinfo {author} {\bibfnamefont {M.}~\bibnamefont {{Halpern}}}, \bibinfo {author} {\bibfnamefont {R.~S.}\ \bibnamefont {{Hill}}}, \bibinfo {author} {\bibfnamefont {N.}~\bibnamefont {{Odegard}}}, \bibinfo {author} {\bibfnamefont {L.}~\bibnamefont {{Page}}}, \bibinfo {author} {\bibfnamefont {K.~M.}\ \bibnamefont {{Smith}}}, \bibinfo {author} {\bibfnamefont {J.~L.}\ \bibnamefont {{Weiland}}}, \bibinfo {author} {\bibfnamefont {B.}~\bibnamefont {{Gold}}}, \bibinfo {author} {\bibfnamefont {N.}~\bibnamefont {{Jarosik}}}, \bibinfo
  {author} {\bibfnamefont {A.}~\bibnamefont {{Kogut}}}, \bibinfo {author} {\bibfnamefont {M.}~\bibnamefont {{Limon}}}, \bibinfo {author} {\bibfnamefont {S.~S.}\ \bibnamefont {{Meyer}}}, \bibinfo {author} {\bibfnamefont {G.~S.}\ \bibnamefont {{Tucker}}}, \bibinfo {author} {\bibfnamefont {E.}~\bibnamefont {{Wollack}}}, \ and\ \bibinfo {author} {\bibfnamefont {E.~L.}\ \bibnamefont {{Wright}}},\ }\href {\doibase 10.1088/0067-0049/208/2/19} {\bibfield  {journal} {\bibinfo  {journal} {\apjs}\ }\textbf {\bibinfo {volume} {208}},\ \bibinfo {eid} {19} (\bibinfo {year} {2013})},\ \Eprint {http://arxiv.org/abs/1212.5226} {arXiv:1212.5226 [astro-ph.CO]} \BibitemShut {NoStop}%
\bibitem [{\citenamefont {{Gehrels}}(1986)}]{Gehrels1986_PoissonCI}%
  \BibitemOpen
  \bibfield  {author} {\bibinfo {author} {\bibfnamefont {N.}~\bibnamefont {{Gehrels}}},\ }\href {\doibase 10.1086/164079} {\bibfield  {journal} {\bibinfo  {journal} {\apj}\ }\textbf {\bibinfo {volume} {303}},\ \bibinfo {pages} {336} (\bibinfo {year} {1986})}\BibitemShut {NoStop}%
\bibitem [{\citenamefont {{G{\"o}{\v{g}}{\"u}{\c{S}} }}\ \emph {et~al.}(1999)\citenamefont {{G{\"o}{\v{g}}{\"u}{\c{S}} }}, \citenamefont {{Woods}}, \citenamefont {{Kouveliotou}}, \citenamefont {{van Paradijs}}, \citenamefont {{Briggs}}, \citenamefont {{Duncan}},\ and\ \citenamefont {{Thompson}}}]{Gogus+1999ApJ_SGR_BurstDistributionStatistics}%
  \BibitemOpen
  \bibfield  {author} {\bibinfo {author} {\bibfnamefont {E.}~\bibnamefont {{G{\"o}{\v{g}}{\"u}{\c{S}} }}}, \bibinfo {author} {\bibfnamefont {P.~M.}\ \bibnamefont {{Woods}}}, \bibinfo {author} {\bibfnamefont {C.}~\bibnamefont {{Kouveliotou}}}, \bibinfo {author} {\bibfnamefont {J.}~\bibnamefont {{van Paradijs}}}, \bibinfo {author} {\bibfnamefont {M.~S.}\ \bibnamefont {{Briggs}}}, \bibinfo {author} {\bibfnamefont {R.~C.}\ \bibnamefont {{Duncan}}}, \ and\ \bibinfo {author} {\bibfnamefont {C.}~\bibnamefont {{Thompson}}},\ }\href {\doibase 10.1086/312380} {\bibfield  {journal} {\bibinfo  {journal} {\apjl}\ }\textbf {\bibinfo {volume} {526}},\ \bibinfo {pages} {L93} (\bibinfo {year} {1999})},\ \Eprint {http://arxiv.org/abs/astro-ph/9910062} {arXiv:astro-ph/9910062 [astro-ph]} \BibitemShut {NoStop}%
\bibitem [{\citenamefont {{ZTF Team}}(2025)}]{ZTF_IPAC_LightCurves}%
  \BibitemOpen
  \bibfield  {author} {\bibinfo {author} {\bibnamefont {{ZTF Team}}},\ }\href {\doibase 10.26131/IRSA598} {\enquote {\bibinfo {title} {Ztf lightcurves},}\ } (\bibinfo {year} {2025})\BibitemShut {NoStop}%
\end{thebibliography}%

\end{document}